\documentclass[twocolumn,secnumarabic,amssymb, nobibnotes, aps, prb]{revtex4-1}
\usepackage{graphicx}
\usepackage{epsfig}
\usepackage{amsmath,amssymb,amsfonts}
\usepackage{psfrag}
\usepackage{enumitem}
\usepackage{color}
\usepackage[colorlinks]{hyperref}
\usepackage{epstopdf}
\usepackage{tikz}
\usepackage{float}

\begin{document}

\title{Superradiant emission stimulated by vortex-antivortex pair production in layered superconductors. }

\author{Alex Gurevich}
\email{gurevich@odu.edu}
\author{Ahmad Sheikhzada}
\email{sheikhzada.ahmad@gmail.com}

\affiliation{Department of Physics, Old Dominion University, Norfolk, VA 23529, USA}

\begin{abstract}

We report numerical simulations of coupled sine-Gordon and heat diffusion equations describing dynamic states stimulated by a trapped vortex driven by dc current in a stack of up to $N=321$ Josephson junctions. It is shown that the Cherenkov wake behind the vortex shuttle trapped in the stack can trigger proliferation of counter-propagating vortices and antivortices which get synchronized and form large-amplitude standing electromagnetic waves. This happens if the dc current density $J$ exceeds a threshold value $J_s$ which can be well below the Josephson interlayer critical current density $J_c$ for  underdamped junctions.  The cavity modes stimulated by the vortex-antivortex pair production cause peaks in the radiated power $P_N(J)$ with a nearly monochromatic spectrum at discrete values of $J$ corresponding to the zero-field Fiske resonances. The power $P_N(J)$ was evaluated for small rectangular stacks in the magneto-dipole approximation and for large stacks in a single mode state. For small stacks, the highest peak in $P_N(J)$ increases rapidly, $P_N\propto N^6$, with the number of junctions at $N\leq 81$ and gradually slows down to $P_N\propto N^2$ at $161\leq N\leq 321$. For stacks larger than the radiated wavelength, we obtained $P_N\propto N^5$ at $N\lesssim 200-300$ and $P_N\propto N^2$ at larger $N$.  For stacks with up to $321$ junctions and representative parameters of Bi$_2$Sr$_2$CaCu$_2$O$_{8+\delta}$, we observed moderate overheating and no hotspots.  The vortex-antivortex pair production can amplify THz radiation from Bi$_2$Sr$_2$CaCu$_2$O$_{8+\delta}$ mesas for which trapping Josephson vortices could be used to stimulate THz emission at subcritical currents and optimize the radiation output. 

\end{abstract}

\maketitle

\section{Introduction}
\label{sec:intro}

The intrinsic Josephson effect \cite{km} in Bi$_2$Sr$_2$CaCu$_2$O$_{8+\delta}$ (Bi-2212)  in which the CuO$_2$ planes sandwiched between the Bi$_2$O$_2$ insulating layers form Josephson junctions (JJs), has caused much interest as a source of coherent radiation from the layered cuprates~   \cite{thz1,thz2,thz3,thz4,thz5}. This effect can be used for the development of compact THz emitters and detectors for fundamental research and applications. Experiments on Bi-2212 mesas have revealed continuous radiation with powers $\sim1 -10^3~ \mu$W at  0.3-10 THz \cite{thz2,thz3,thz4,thz5} and the appearance of hotspots in the mesas \cite{ths0,ths1,ths2,ths3,ths4,ths5,ths6}. Such THz emitters can operate both at 4.2 K and 77 K~ \cite{thN1,thN2}.

The THz radiation from the JJ stack results from the excitation of collective modes of synchronized JJs ~\cite{thz1,thz2,thz3,thz4}. Such modes can produce  oscillating charge density at the edges of the stack and can be enhanced by electromagnetic coupling with surrounding structures ~\cite{thz1,thz2,thz3,thz4} or mesa arrays ~\cite{coupl1,coupl2}. Reaching higher radiation powers $P_N$ requires better synchronization of JJs so that $P_N$ would increase quadratically with the large number $N$ of CuO planes in the crystal in a superradiant state. To improve the synchronization of JJs in Bi-2212 mesas it was proposed to use traveling electromagnetic (EM) waves \cite{kras1,kras2,kras3}, periodic inhomogeneity of the interlayer current density \cite{inh1,inh2}, in-plane magnetic fields \cite{rah} or currents \cite{inpls}. 
 
In this paper we consider a mechanism of intrinsic synchronization of JJs and radiation stimulated by Cherenkov instability of a current-driven Josephson (J) vortex trapped in the stack ~\cite{sg}. Cherenkov radiation produced by moving vortex structures in JJ stacks is usually regarded as an extra contribution to the total radiation output ~ \cite{thz2,cher}.  Here we consider a nonlinear Cherenkov synchronization of JJs stimulated by a single vortex which triggers spontaneous production of vortex-antivortex (V-AV) pairs above a threshold current $I_s$ in the absence of dc magnetic field. Such V-AV pair production has been observed in numerical simulation of single JJs ~\cite{aj1,aj2}, two and three-stacked JJs ~\cite{a1,a2,a3,a4}, annular JJs ~\cite{annul},  JJ arrays ~\cite{jja1,jja2}, Josephson multilayers ~\cite{sg} and other systems described by coupled sine-Gordon equations ~\cite{sg1,sg2}.  This effect is most pronounced in underdamped JJ stacks in which the threshold current $I_{s}$  can be well below the interlayer Josephson critical current $I_c$. 

At $I<I_{s}$ a single vortex trapped in an underdamped JJ stack bounces back and forth turning into antivortex upon each reflection from the edges of the stack \cite{sg}. At $I>I_{s}$ such V-AV shuttle triggers proliferation of V-AV pairs which get synchronized and form a large-amplitude standing EM wave. Both the amplitude and the frequency of such resonant mode increase by orders of magnitude as compared to those of the V-AV shuttle at $I<I_{s}$.  This mode causes oscillations of the magnetic moment and magneto-dipole radiation from small JJ stacks with the emitted power $P_N\propto N^6$ increasing rapidly with the number of JJs, as was shown by numerical simulations of up to $81$ JJs ~\cite{sg}. Such strong increase of $P_N$ with $N$ mostly results from the increase of the number of produced V-AV pairs and the magnetic flux of a vortex as the thickness of the stack increases.  Extending these results to bigger stacks with $N\gtrsim 10^2-10^3$ typical of Bi-2212 mesas requires extensive numerical simulations to address the following questions: 1. How far can the strong increase of $P_N\propto N^6$ with $N$ continue and at what $N$ could $P_N$ eventually level off or start decreasing? 2. How much is $P_N$ limited by overheating which was  disregarded in Ref. \onlinecite{sg}. 3. How can the emission and its angular distribution change in mesas with lateral dimensions greater than the radiated EM wavelength $\sim 10^{-1}$ mm at which the magneto-dipole approximation is not applicable? In this work we address these issues by calculating $P_N$ and the radiation spectrum for JJ stacks with up to 321 junctions. 

The paper is organized as follows.  In Sec. II we present the coupled sine-Gordon equations solved self-consistently with an equation for the mean temperature $T$ of the JJ stack. A qualitative picture of a standing EM wave stimulated by a V-AV shuttle is given in Sec. III. Calculations of the magneto-dipole radiated power $P_N(\beta)$ as functions of the driving dc current and the number of layers for a single vortex trapped in the central JJ of a small stack are given in Sec. IV. In Sec V we calculate $P_N(\beta)$ for one trapped vortex in each JJs. In Sec. VI we evaluate the radiated power caused by a resonant  mode in a rectangular JJ stack of arbitrary dimensions. Sec. VII concludes with a discussion of the results.

\section{Main Equations}
\label{eqs}
 
\begin{figure}[h!]
\includegraphics[width=\columnwidth]{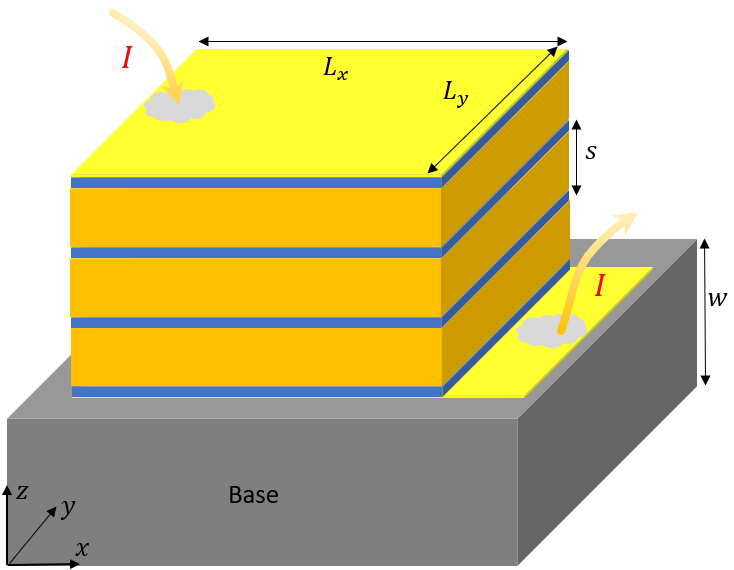}
\caption{A stack of intrinsic Josephson junctions (brown) between superconducting layers. A dc current $I$ is injected from the top layer.}
\label{mesa}
\end{figure}

In this work we solve numerically generic sine-Gordon equations  \cite{km,thz1,thz2,thz3,thz4,sakai,klein,bul,kleiner,machida,tachiki,lin-sust} which  
describe dynamic states of a stack of $N$ coupled JJs shown in Fig. \ref{mesa}:
\begin{gather}
\theta_n''= \left[1-\zeta(T)\Delta_d\right]\left[\alpha_J(T)\sin\theta_n+\eta\dot{\theta_n}+\ddot{\theta_n}\right],
\label{phase} \\
b_n = \left[1-\zeta(T)\Delta_d\right]^{-1}\theta_n'.
\label{Bn}
\end{gather}
Here $\theta_n(x,t)$  is the phase difference across the $n$-th junction, $b_n=B_n(x,t)/B_0$ the dimensionless magnetic field parallel to the layers, $B_0=\mu_0J_c\lambda_{c0}=\phi_0/2\pi s\lambda_{c0}$,  $\lambda_c$ and $\lambda$ are the penetration depths of the parallel magnetic field ${\bf B}$ along and across the layers, respectively, $\lambda_{c0}=\lambda_c(T_0)$, $s$ is the spacing between the layers, $\phi_0$ is the magnetic flux quantum, $\Delta_d f_n\equiv f_{n+1}+f_{n-1}-2f_n$ is the lattice Laplacian, $J_c$ is the Josephson critical current density,  $\eta=\sigma_c\lambda_{c0}/c\epsilon_0\sqrt{\epsilon_c}$ is a damping parameter,  $\sigma_c$ is the interlayer quasiparticle conductivity, $\epsilon_0$ and $\mu_0$ are the vacuum electric and magnetic permittivities, the prime and overdot denote partial derivatives with respect to the dimensionless coordinate $x/\lambda_{c0}$ and time $t \omega_{J0}$, respectively, $\omega_{J0}=c/\sqrt{\epsilon_c} \lambda_{c0}$ is the Josephson plasma frequency at $T=T_0$, $c$ is the speed of light and $\epsilon_c$ is the dielectric constant along the crystal $c$ axis. We consider 1d solutions $\theta_n(x,t)$ and $b_n(x,t)$ independent of $y$ and disregard the effects of charge imbalance ~\cite{lin-sust} and in-plane quasiparticle currents ~\cite{artem,thz5} in Eqs. (\ref{phase})-(\ref{Bn}).   

At the edges of the stack we imposed the boundary condition $\theta'_n(0)=\theta'_n(L_x)=0$ of zero in-plane super current density. Other boundary conditions account for a dimensionless dc uniform current density $\beta=J/J_c(T_0)$  injected through the top JJ $(n=1)$ and collected from the bottom JJ $(n=N)$ ~\cite{kleiner,machida,tachiki}. These boundary conditions were incorporated in the matrix Eqs. (\ref{phase})-(\ref{Bn}) which were solved numerically using the method of lines ~\cite{mdln,mdabm} as described in Appendix \ref{A}.

The parameters $\alpha_J(T)$ and $\zeta(T)$ take account of dependencies of $J_c(T)$ and $\lambda(T)$ on the JJ temperature $T(J)$, 
\begin{gather}
\alpha_J=\frac{J_c(T)}{J_c(T_0)}, 
\qquad
\zeta=\frac{\lambda^2(T)}{s^2}, 
\label{par}
\end{gather}
where $\zeta(T)$ quantifies the inductive coupling of the layers. We use $J_c(T)$ for SIS junctions and an interpolation formula for $\lambda(T)$ in Bi-2212 single crystals~ \cite{ths3,lambda}:  
\begin{gather}
J_c(T)/J_c(0)= 1-(T/T_c)^2,
\label{Jct} \\
\lambda^2(0)/\lambda^2(T)=\left[1-(T/T_c)^6\right]\left(1-0.6T/T_c\right).
\label{lt}
\end{gather} 
A mean temperature of the JJ stack $T(J)$ is calculated self-consistently by solving Eqs. (\ref{phase})-(\ref{Bn}) and a stationary heat diffusion equation. As shown in Appendix \ref{B}, $T(J)$ in a thin JJ stack of thickness $d$ on a base of thickness $w\gg d$ is determined by the equation: 
\begin{gather}
\frac{1}{w}\int_{T_0}^T\kappa_c(T)dT= \label{temp} \\
\frac{\hbar J_{c0}}{2et_aL_x}\sum_{n=1}^N\int_{t_0}^{t_0+t_a} dt\int_0^{L_x} [\eta\dot \theta_{n}^2+\alpha_J\dot{\theta}_n\sin\theta_n]dx,
\nonumber
\end{gather} 
where $\kappa_c(T)$ is the thermal conductivity of the base along the c-axis. The right hand side of Eq. (\ref{temp}) is the power generated per unit area of the stack. In our simulations $t_0=2000$ was chosen to be larger than the transient time after which the steady-state solutions $\theta_n(t)$ set in and $t_a=200$ is an averaging time to calculate steady-state $T(\beta)$ and the radiation power.  
 
Simulations started at a small current $\beta\ll 1$ and run until a steady state $\theta_n(t)$ was reached at $t>t_0$. Then these solutions were used to calculate $T(\beta)$ in Eq. (\ref{temp}) and update $\zeta(T)$ and $\alpha_J(T)$ in Eqs. (\ref{phase})-(\ref{Bn}) for the next step $\beta+\Delta\beta$, where $\Delta\beta\leq 0.01$. In this way Eqs. (\ref{phase})-(\ref{temp}) were solved iteratively for a sequence of driving currents $\beta_{i+1}=\beta_i+\Delta\beta$ using  $\theta_n(t,\beta_i)$ obtained at the preceding $i$-th step as initial conditions for Eqs. (\ref{phase})-(\ref{Bn}) at the $i+1$-th step and to update $T(\beta_i)$ in Eq. (\ref{temp}). For the input material parameters, we took $T_c=85$ K, $\epsilon_c=12$, $s=1.5$ nm, $J_c(T_0)=200$ A/cm$^2$,  $\lambda(0)=260$ nm,  $\lambda(T_0)=264$ nm in a Bi-2212 crystal at the ambient temperature $T_0=4.2$ K, $\lambda_{c0}=295\,\mu$m, $\omega_{J0}=0.29$ THz, $L_x=\lambda_{c0}$ and the base thickness $w=30\,\mu$m. The temperature dependence of $\eta(T)$ was disregarded and $\kappa_c(T)=\kappa_0(T/T_0)^a$ with $\kappa_0=0.32$ W/mK and $a=0.67$ ~\cite{ths3} was used.

\section{Resonant modes stimulated by a vortex shuttle}
\label{chv}

We start with an overview of the physics of the V-AV pair production by a moving vortex trapped in a stack with 21 JJs. This case was considered in Ref. \onlinecite{sg} in a model with a constant dc bias current $\beta$ through each JJ added in Eq. (\ref{phase})~ \cite{lin-sust}. Here we adopt a more consistent approach in which $\beta$ is taken into account through the boundary conditions at the top and the bottom JJ ~\cite{kleiner,machida,tachiki}. Both approaches produce the same qualitative results. 

\begin{figure}[h!]
\includegraphics[width=\columnwidth]{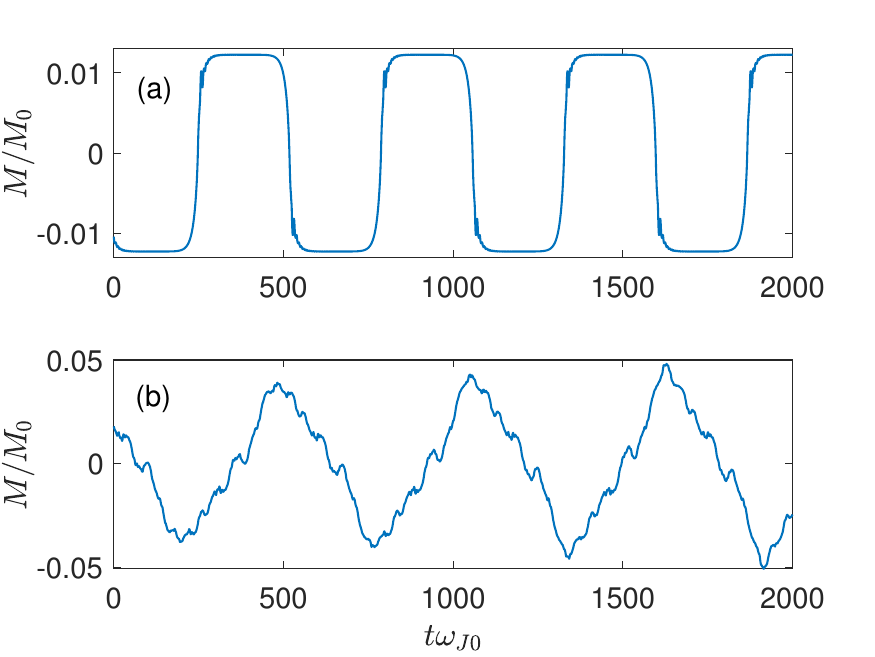}
\caption{Temporal oscillations of a magnetic moment $M(t)$ due to reflections of vortices and antivortices from the edges at $\eta=0.1$. (a) $M(t)$ caused by a single vortex trapped in a central JJ at $\beta=0.585<\beta_s$, (b) $M(t)$ caused by a bouncing flux structure with one vortex per layer at $\beta=0.52<\beta_s$. }
\label{shuttle}
\end{figure}

Dynamics of a vortex in the current-biassed JJ stack depends crucially on the damping parameter $\eta$. If $\eta\gtrsim 1$ a vortex initially trapped in a JJ is pushed by current toward the edge of the stack and exits. In an underdamped stack with $\eta\ll 1$ the vortex colliding with the edge changes polarity and gets reflected as antivortex. This process can also be viewed as a vortex exit followed by penetration of antivortex. The polarity change upon each reflection of the vortex from the edges results in a vortex-antivortex (V-AV) shuttle and temporal oscillations of the magnetic moment $M(t)$ with the flight frequency $f_v\simeq v/2L_x$ depending on the vortex velocity $v$ and the JJ length $L_x$. Such oscillations of $M(t)$ calculated from Eq. (\ref{M}) are shown in Fig. \ref{shuttle} for a vortex moving along a central JJ (a) and one trapped vortex per each JJ (b). In both cases we have $f_v\sim 10^{-2}\omega_{J0}$. The ripples on $M(t)$ seen in Fig. \ref{shuttle} result from the bremsstrahlung caused by acceleration of the vortex due to its attraction to the AV image at the edges.  

\begin{figure}[h!]
\includegraphics[width=\columnwidth]{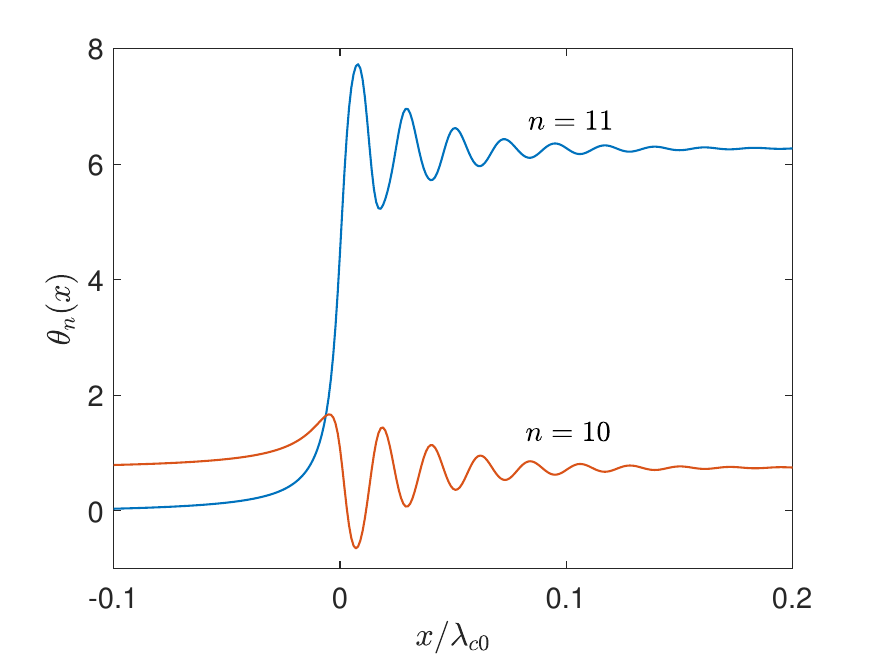}
\caption{Phase profile of a vortex propagating along  a long 
central junction ($n = 11, L_x\gg \lambda_{c0}$) and the trailing tail of Cherenkov radiation on the neighboring JJ ($n = 10$) calculated from Eq. (\ref{phase}) at $\eta=0.1$ and $\beta=0.685$.}
\label{fig3}
\end{figure}

\begin{figure}[h!]
\includegraphics[width=\columnwidth]{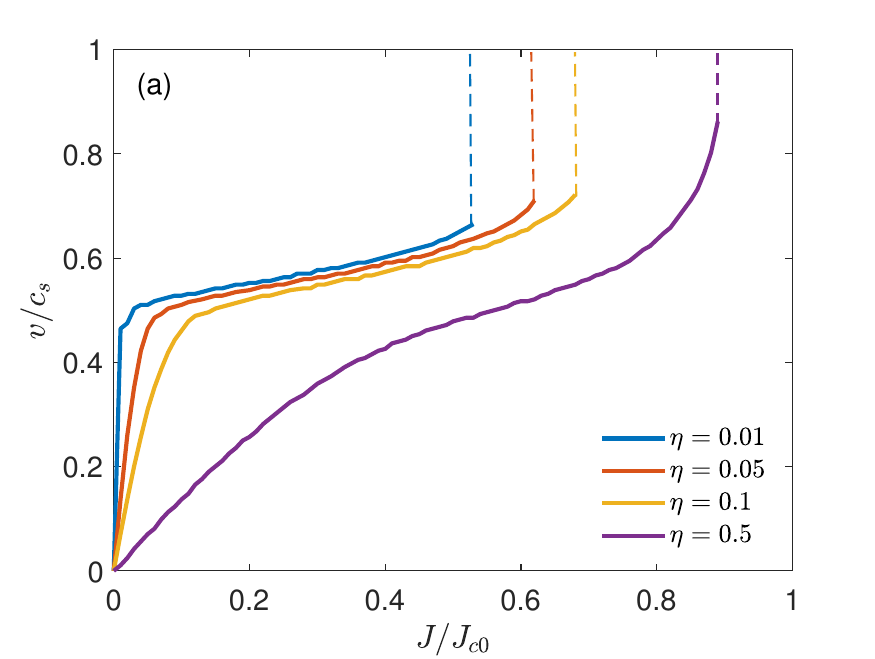}\\
\includegraphics[width=\columnwidth]{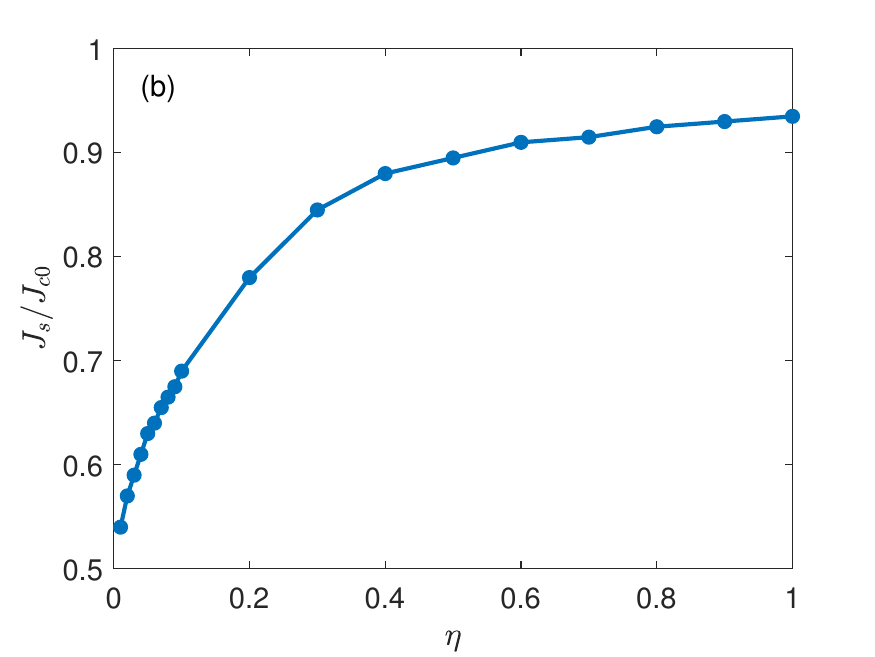}
\caption{(a) Stationary velocities of a vortex moving along the central JJ as a function of the bias current at different damping parameters $\eta$.  (b) The threshold current density $J_s$ corresponding to the endpoints of the $v(\beta)$ curves in (a) 
as a function of $\eta$ calculated for $\zeta=30962$.}
\label{fig4}
\end{figure}

\begin{figure}[h!]
\includegraphics[width=\columnwidth]{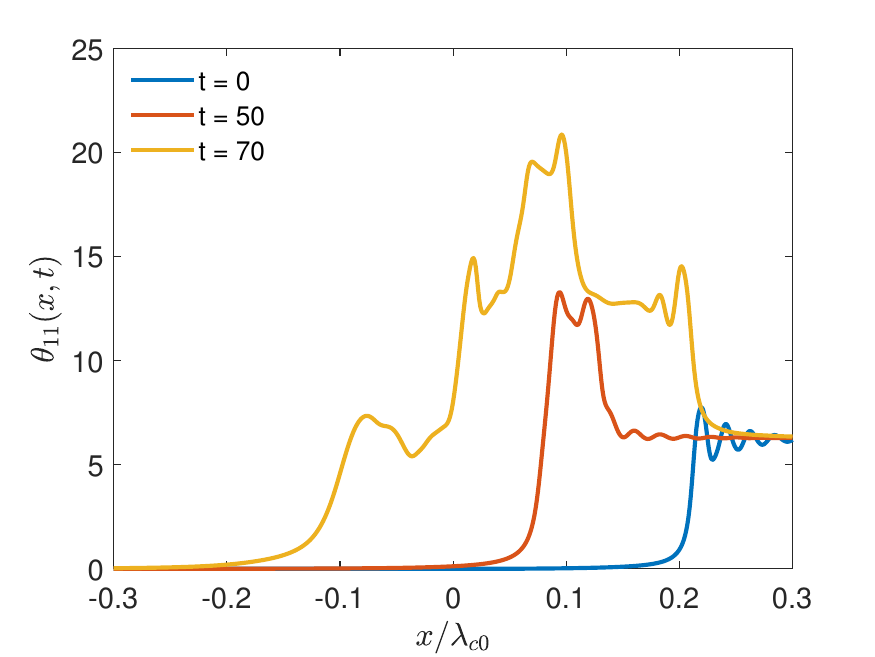}
\caption{Snapshots of V-AV pairs generated by a vortex propagating along the central junction in a long stack $(L_x\gg \lambda_c)$ calculated at 
$\eta=0.1$ and $\beta=0.69$.}
\label{fig5}
\end{figure}


The velocity of V-AV shuttle is controlled by the balance of the Lorentz force and quasiparticle viscous drag at small currents and by the  
radiation losses at higher currents ~\cite{thz2}. The radiation losses are caused by the trailing Cherenkov wake behind a moving vortex shown in Fig. \ref{fig3}. To see interplay of these mechanisms, consider a vortex propagating uniformly along a long JJ stack. Shown in Fig. \ref{fig4} is the velocity $v(\beta)$ of a vortex driven along the central JJ calculated from Eq. (\ref{phase}) without thermal feedback, where $v(\beta)$ is normalized to $c_s=cs/\lambda\sqrt{\epsilon_c}$ of the order of the Swihart velocity ~\cite{thz2}. The initial steep increase of $v(\beta)$ controlled by the weak viscous drag is followed by the sharp decrease in the slope of $v(\beta)$ due to radiation losses.   At $\eta=0.1$ the radiation-limited velocity $v\simeq 0.5c_s$ determines the flight oscillation frequency $f_v=v/2L_x$ of bouncing vortex. For $L_x=\lambda_{c0}$ used in our simulations,  $f_v\sim (s/\lambda)\omega_{J0}\sim 10^{-2}\omega_{J0}$, consistent with Fig. \ref{shuttle}.
 
The radiation-controlled parts of $v(\beta)$ in Fig. \ref{fig4} are terminated at the end points $\beta=\beta_s$. At $\beta>\beta_s$ the steady-state propagation of the vortex becomes impossible because the Cherenkov wake gets so strong that it causes spontaneous production of V-AV pairs. 
This process is illustrated by Figs. \ref{fig3} and \ref{fig5}, which show a Cherenkov wake   
with $5\pi/2<\theta_n<7\pi/2$ behind the moving vortex. A uniform state with $5\pi/2<\theta_n<7\pi/2$ is unstable with respect to small perturbations $\delta\theta_n\ll 1$ but gets stabilized in a domain of finite length.  As $\beta$ increases, the amplitude
and the width of this domain grow so that at $\beta=\beta_s$ it becomes unstable and  
splits, triggering a cascade of expanding V-AV pairs, as shown in Fig \ref{fig5}. In turn, the Cherenkov wake on the central JJ induces new V-AV pairs on the neighboring JJs. Those V-AV pairs start splitting and propagating both along and across the stack, vortices and antivortices bundle in spatially-separated multi-quanta flux spots (macrovortices)~\cite{sg} and counter-propagating flux spots with opposite polarity ~\cite{supp}.  According to Fig. \ref{fig4}(b), the threshold current density $J_s(\eta)$ for the V-AV pair production in underdamped JJ stacks can be well below the interlayer $J_c$.   

\begin{figure}[h!]
\includegraphics[width=\columnwidth ]{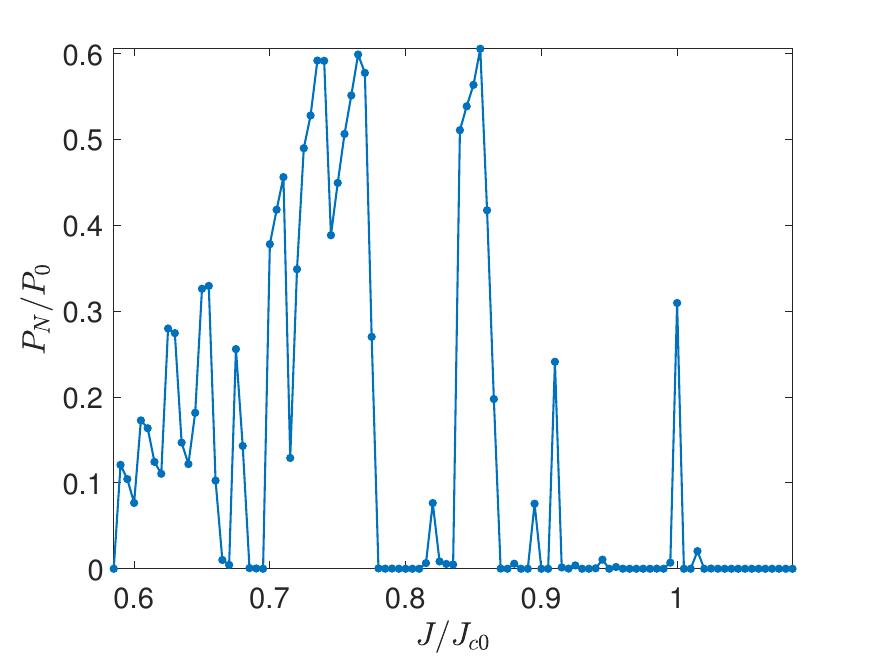}
\includegraphics[width=\columnwidth ]{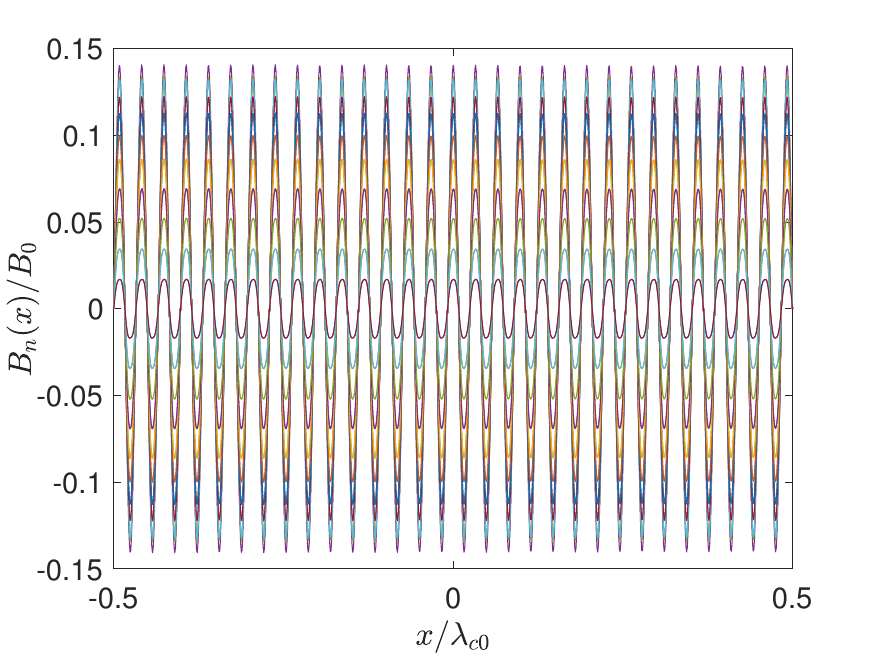}
\caption{(top) The radiated power, $P_N(\beta)$ as a function of current calculated for a stack of length $L_x=\lambda_{c0}$ and $N=21$.
(bottom): Snapshots of a resonant mode in $B_n(x,t)$ on different JJs at the peak in 
$P_N(\beta)$ above at $\beta=0.85$.}
\label{P21}
\end{figure}

Flux spots produced by the V-AV shuttle collide with the edges of the stack and change polarity upon each reflection, as shown in Ref. \onlinecite{supp}.  After repeated bouncing back and forth, these flux spots get synchronized and form a standing EM wave which can cause temporal oscillations of the magnetic moment of the stack:
\begin{equation}
M(t)=M_0\sum_n \int_0^{L_x/\lambda_{c0}} b_n(x)dx,
\label{M}
\end{equation}
where $M_0=B_0 s \lambda_c L_y/\mu_0=\phi_0 L_y/2\pi\mu_0$. The oscillations of $M(t)$ result in radiation from JJ stacks. For stacks much smaller than the EM wavelength, the radiated power  can be evaluated in magneto-dipole approximation $P_N=\mu_0\langle\ddot{M}^2\rangle/6\pi c^3$, where $\langle ...\rangle$ denotes time averaging ~\cite{jack}. It is convenient to write $P_N$ in the form;
\begin{gather}
P_N=P_0G_N,\qquad P_0=\frac{c\phi_0^2 L_y^2}{24\pi^3\mu_0\epsilon_c^2\lambda_c^4},
\label{P} \\
G_N=\int_{t_0}^{t_0+t_a}\!\bigg[\sum_{n=1}^N \int_0^{L_x/\lambda_{c0}} \ddot{b}_n(x,t)dx\bigg]^2\frac{dt}{t_a}.
\label{GN}
\end{gather}
Here $t_0=2000$ is chosen to be much larger than a transient time to reach a steady-state, $t_a=200$ is an averaging time and  
$G_N$ accounts for all harmonics of $B_n(x,t)$. For $L_y=\lambda_c=295~\mu$m and $\epsilon_c=12$, we have $P_0\simeq 10^{-13}$ W.  Figure \ref{P21} shows the calculated $P_N(\beta)$ which has multiple peaks diminishing  at $\beta\agt 1$. The peaks in $P_N(\beta)$ occur at currents for which the resonant mode $B_n(x,t)$ has odd numbers of half-periods along the stack.  

The standing EM wave shown in Fig. \ref{P21} is formed by bundles of alternating J vortices and antivortices. The magnitude of $M(t)$ is proportional to the magnetic flux $\Phi$ of these bundles, both $\Phi$ and $P_N$ increasing strongly with the number $N$ of JJs in a stack with $d=sN<2\lambda$. This is because the flux of a J vortex $\phi$ in a thin stack can be much smaller than $\phi_0$ ~\cite{sg}, similar to the well-known result for the Abrikosov vortex in a thin film ~\cite{vf0,vf1,vf2,vf3}:   
\begin{equation}
\phi(u) = \phi_0\left[1-\frac{\cosh(u/\lambda)}{\cosh(d/2\lambda)}\right].
\label{ph}
\end{equation} 
Here $u$ is the distance of the vortex from the center of the stack. The flux $\phi(u)$ decreases with $u$ and vanishes at the surfaces $u=\pm d/2$, where the vortex is extinguished by its AV image ~\cite{vf1,vf2}. If $d\ll 2\lambda$, Eq. (\ref{ph}) gives:
\begin{equation}
\phi(u)\simeq \frac{\phi_0 N^2}{8}\left(\frac{s}{\lambda}\right)^2\!\left(1-\frac{4u^2}{d^2}\right),\quad N\lesssim N_c=\frac{2\lambda}{s}.
\label{phi}
\end{equation}
The reduction of $\phi(u)$ in a thin stack results from the confinement of vortex currents in a small area $\sim d^2\ll \lambda^2$,  which can also be interpreted in terms of partial extinguishing of the vortex field by its AV images.  
For $N=21$, $s=1.5$ nm and $\lambda=264$ nm, we get $\phi(0)\simeq 1.8\cdot 10^{-3}\phi_0$.  At $N\lesssim N_c $ the flux $\phi$ is greatly reduced but 
$\Phi\sim \phi_0(N/N_c)^2N$ of flux spots and $P_N\propto N^6$ increase rapidly with $N$.  For $\lambda(T_0)=264$ nm and $s=1.5$ nm, we obtain $N_c\simeq 352$ at $4.2$ K. Overheating increases $N_c(T)$, for instance, at $T=50$ K, Eq. (\ref{lt}) yields $N_c(T)\simeq 440$. 

\section{Single vortex trapped in the central junction}
\label{sv}

\begin{figure*}[ht]
\includegraphics[width=\columnwidth ]{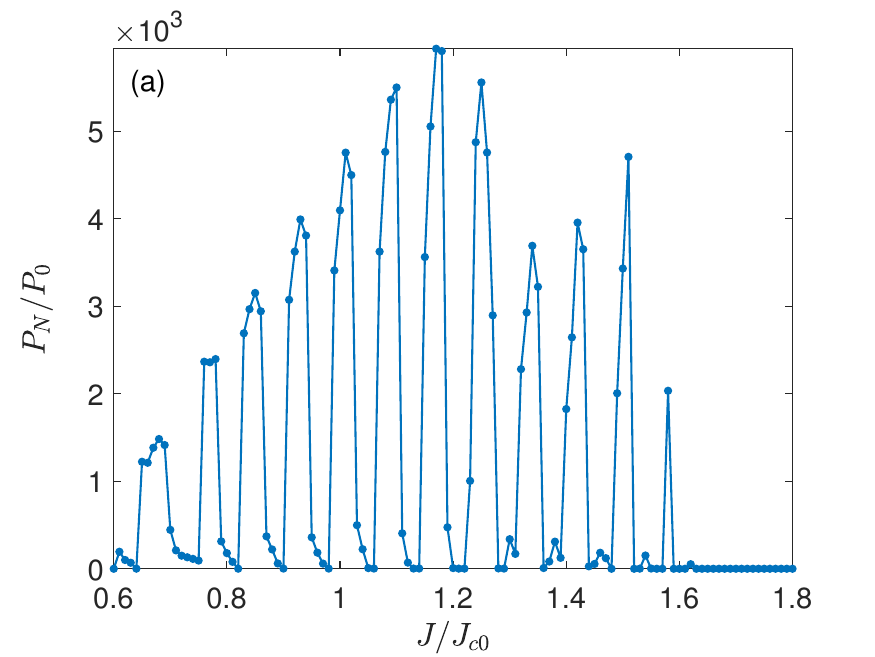}
\includegraphics[width=\columnwidth ]{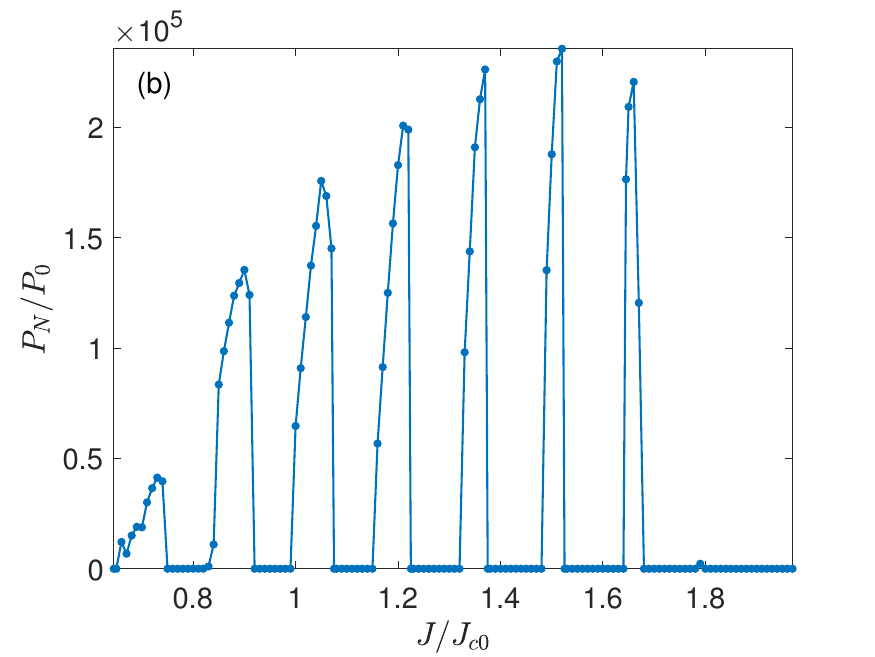}
\includegraphics[width=\columnwidth ]{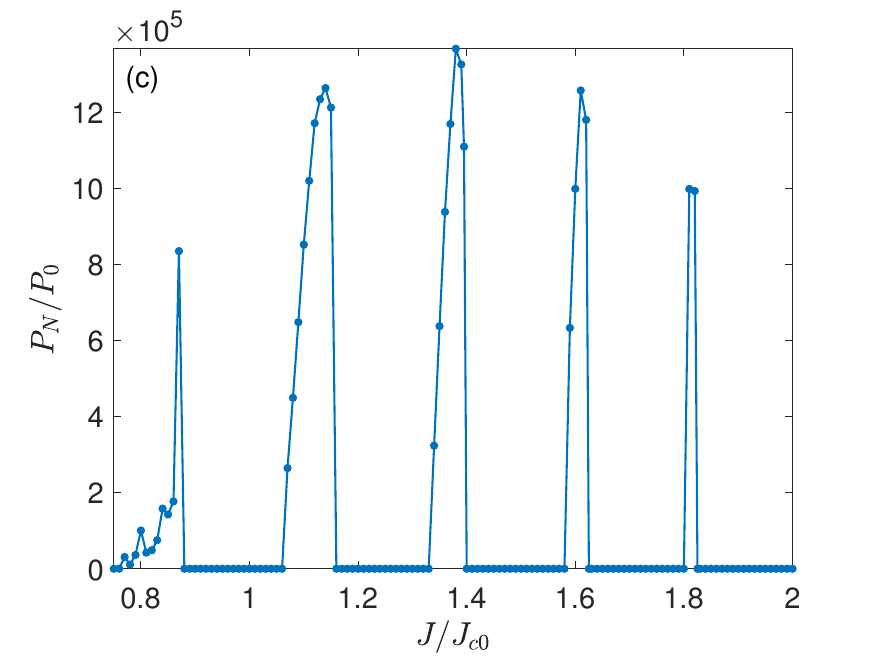}
\includegraphics[width=\columnwidth ]{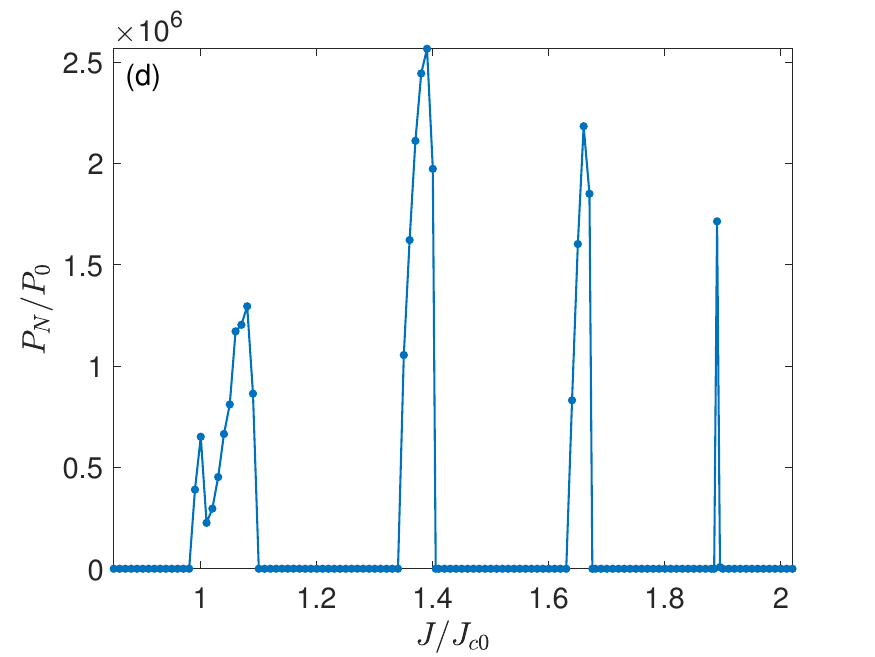}
\caption{The radiation power, $P_N(\beta)$ calculated for a stack with $81$, $161$, $261$ and $321$ JJs as functions of the bias current.}
\label{Psv}
\end{figure*}

\begin{figure}[h!]
\includegraphics[width=\columnwidth ]{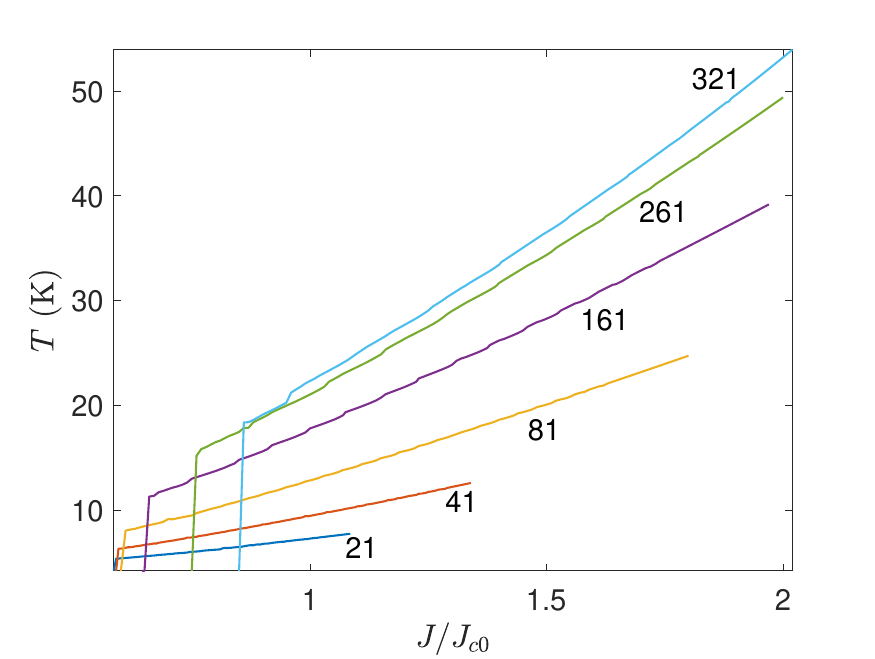}
\caption{ The mean temperatures $T(J)$ of JJ stacks with different number of layers. The sharp increase of $T(J)$ occurs at the onset of the V-AV pair production. }
\label{Tsv}
\end{figure}

\begin{figure}
\includegraphics[width=\columnwidth ]{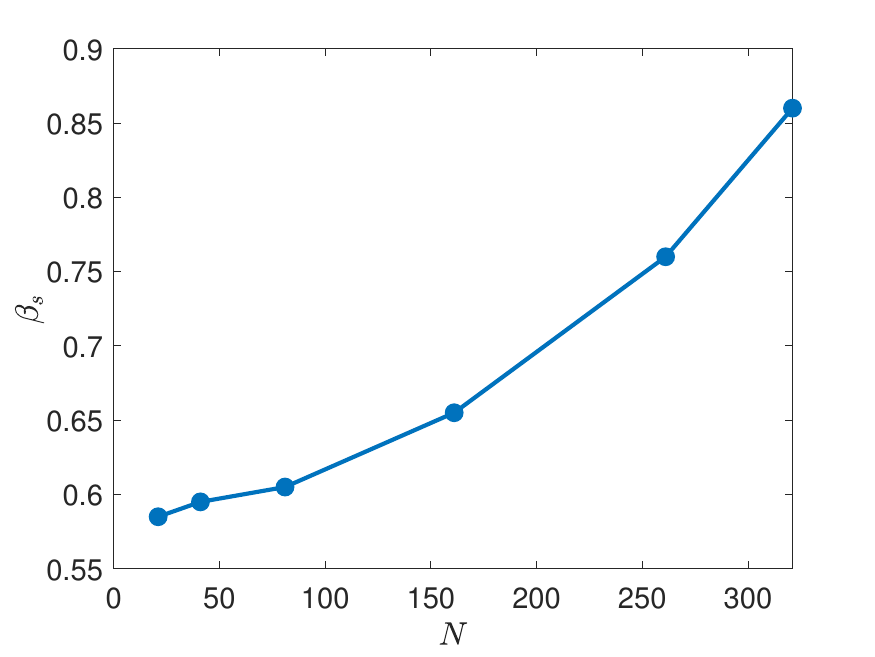}
\caption{ The current threshold of the V-AV pair production $\beta_s$ as a functions of $N$. }
\label{Jsv}
\end{figure}

In this section we present numerical results for a dynamic state stimulated by a V-AV shuttle trapped  
in the central JJ of a stack with $\eta=0.1$~\cite{lin-sust} and $21 \leq N\leq  321$.  After the standing EM wave sets in at $\beta>\beta_s$, the steady-state $P_N(\beta)$ and $T_N(\beta)$ were calculated. The resulting $P_N(\beta)$ evaluated in the magneto-dipole approximation and $T_N(\beta)$  raise sharply at the current onset of the V-AV pair production $J_s$, as shown in Figs. \ref{Psv} and \ref{Tsv}. Here $J_s(N)$ increases with $N$ but remains below the interlayer $J_c$, as shown in Fig. \ref{Jsv}. 

\begin{figure}[h!]
\includegraphics[width=\columnwidth ]{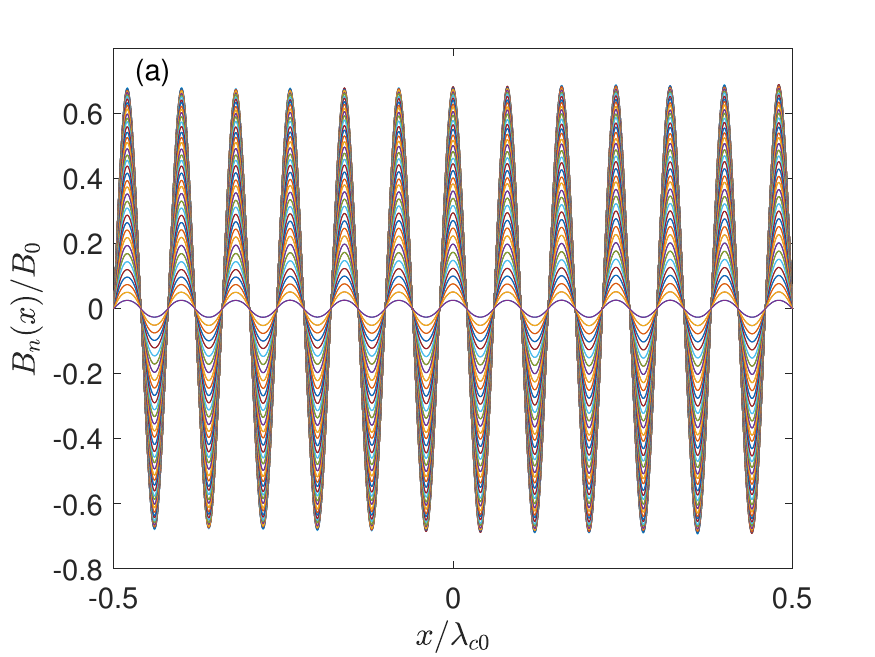}
\includegraphics[width=\columnwidth ]{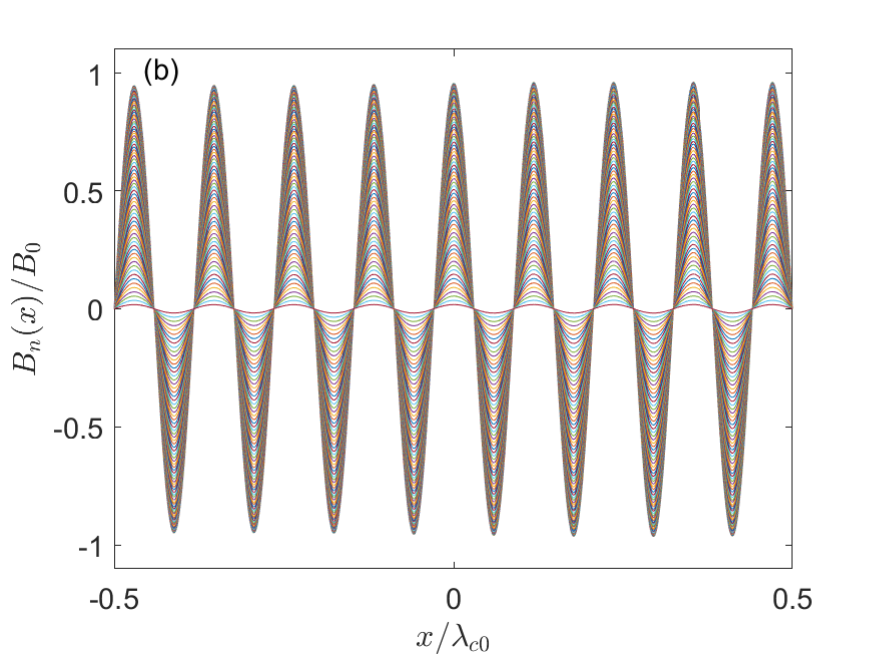}
\includegraphics[width=\columnwidth ]{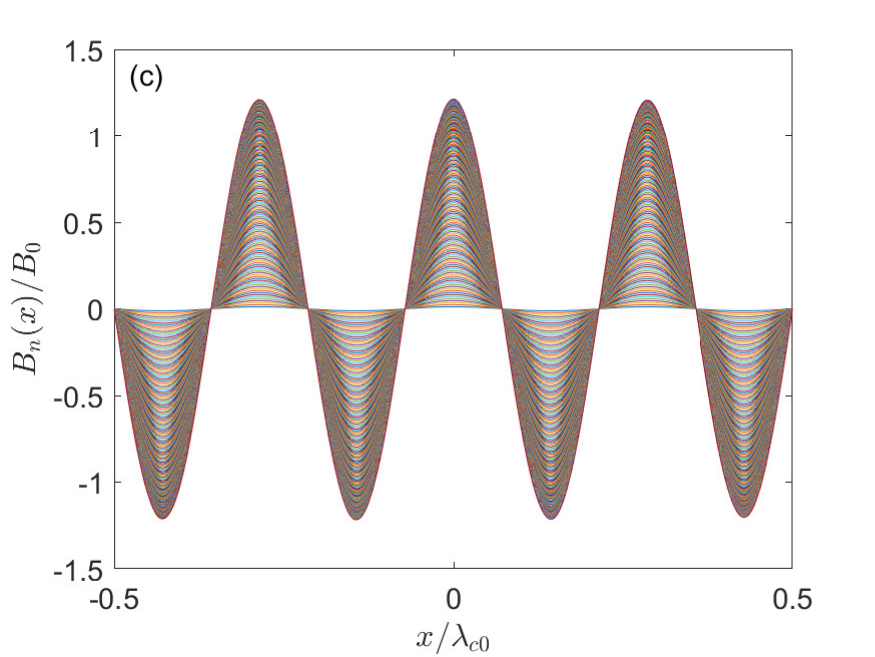}
\caption{Snapshots of odd resonant modes in $B_n(x,t)$ on different
layers in a stack with: (a) $N=81$ at a radiation peak at $\beta=1.17$, (b)  $N=161$ at a radiation peak at $\beta=1.52$, (c) $N=321$ at the radiation peak at $\beta=1.39$.}
\label{Bsv}
\end{figure}

\begin{figure}[h!]
\includegraphics[width=\columnwidth ]{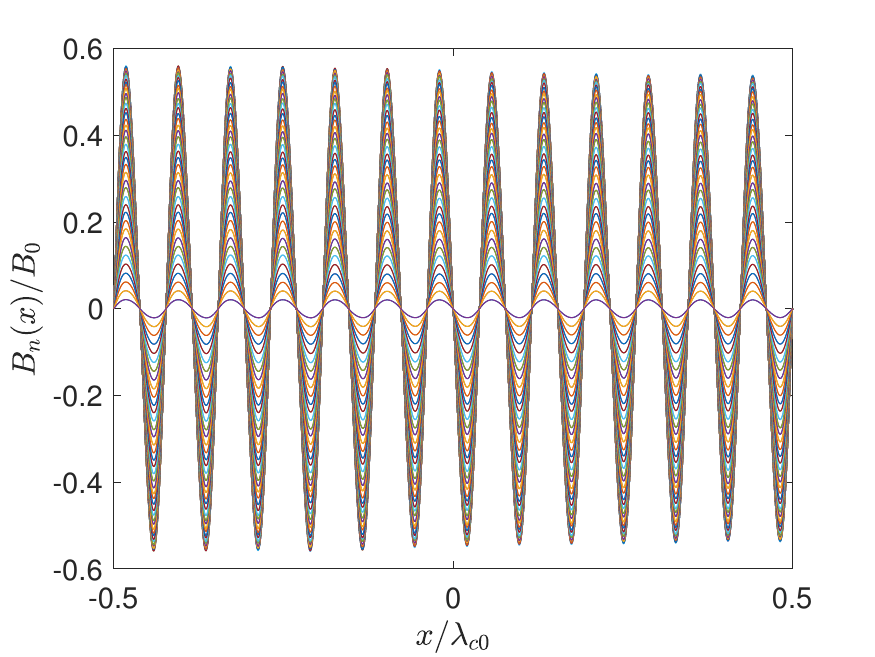}
\caption{A snapshot of even mode $B_n(x,t)$ on different
layers at minimum $P_N(\beta)$ at $\beta=1.2$ and $N=81$. }
\label{Bmin}
\end{figure}

The radiated power $P_N(\beta)$ as a function of current has sharp peaks, whereas $T(\beta)$ increases smoothly with $\beta$ after the initial jump at $\beta=\beta_s$.  The peaks in $P_N(\beta)$ occur for standing waves in $B_n(x,t)$ with odd numbers $m$ of half-periods, as shown in Fig. \ref{Bsv}. In turn, the minima in $P_N(\beta)$ correspond to even numbers $m$ of half-periods of $B_n(x,t)$ with about the same amplitudes as for the odd modes (see Fig. \ref{Bmin}).  The unequal number of positive and negative peaks in $B_n(x,t)$ with odd $m$ causes temporal oscillations of $M(t)$ proportional to the areas of the single peak. The field amplitudes of the strongest resonant mode increase from  $0.14B_0$ at $N=21$ to $1.2B_0$ at $N=321$, where $B_0=\phi_0/2\pi s\lambda_{c0}$ is much larger than the lower critical field $B_{c1}=(\phi_0/4\pi\lambda\lambda_c)[\ln(\lambda/s)+1.12]$ ~\cite{clem}. The wavelength of the resonant mode at the highest peak in $P_N(\beta)$ increases with $N$, so $M_N(\beta)$ proportional to the area under the peak in $B_n(x,t)$ increases strongly with $N$, whereas the heights of peaks in $P_N(\beta)$ first increase with $\beta$ and then decrease with $\beta$.  For  
$L_y=L_x=\lambda_{c0}=295~\mu$m, $\epsilon_c=12$ and $P_0\simeq 10^{-13}$ W, we have $\tilde{P}_{321}\simeq 0.25 ~\mu$W for a stack with $N=321$ at $\beta\approx 1.4$ according to Fig. \ref{Psv}(d).  

\subsection{Spectral analysis and resonant frequencies.}

\begin{figure}[h!]
\includegraphics[width=\columnwidth ]{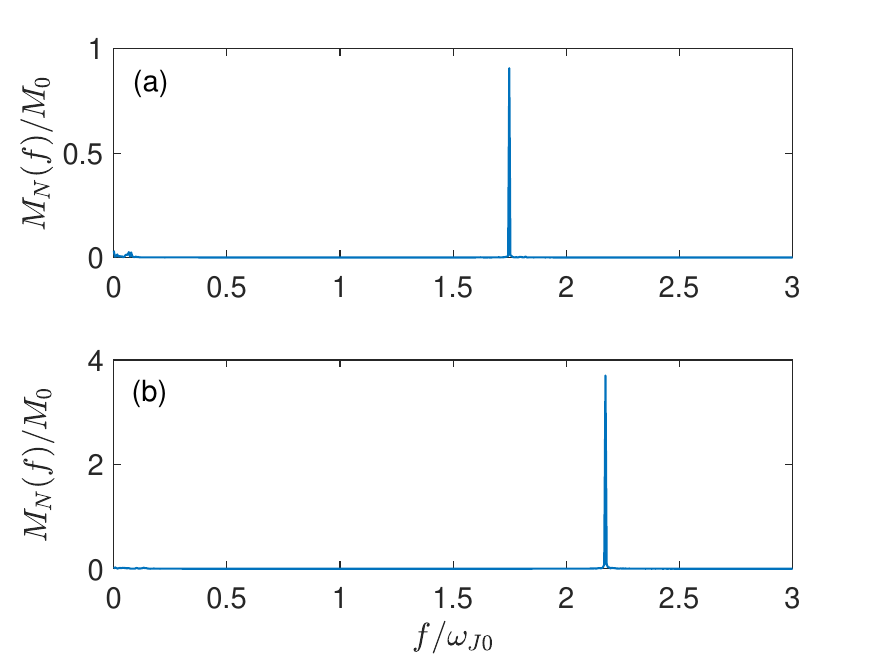}
\caption{Fourier spectra of $M(t)$ at the highest peaks in $P_N(\beta)$: (a) $N=81$ at $\beta=1.17$; (b) $N=161$ at $\beta=1.52$. }
\label{Msv}
\end{figure}

The spectral analysis of $M_N(\beta,t)$ shows that the standing waves producing the peaks of $P_N(\beta,t)$ are practically monochromatic, the amplitudes of fundamental harmonics increase strongly with $N$. Shown in Fig. \ref{Msv} are the Fourier spectra of $\tilde{M}_N(t,\beta)$  for the stacks with $N=81$ and $161$, where the tilde marks $M_N(t,\beta)$ and $P_N(\beta)$ corresponding to the highest peak in $P_N(\beta)$ with frequencies $\tilde{f}_N\simeq (1-2.2)\omega_{J0}$.  The Fourier spectra of $\tilde{M}_N(t,\beta)$ also contain small low-frequency harmonics and the third harmonic with amplitude $\sim 10^{-3}$ of that of the main harmonic. Contributions of the low-frequency and high-frequency harmonics to $P_N(\beta)$ are negligible. 

\begin{figure}[h!]
\includegraphics[scale =0.41,trim={40 210 20 80mm},clip ]{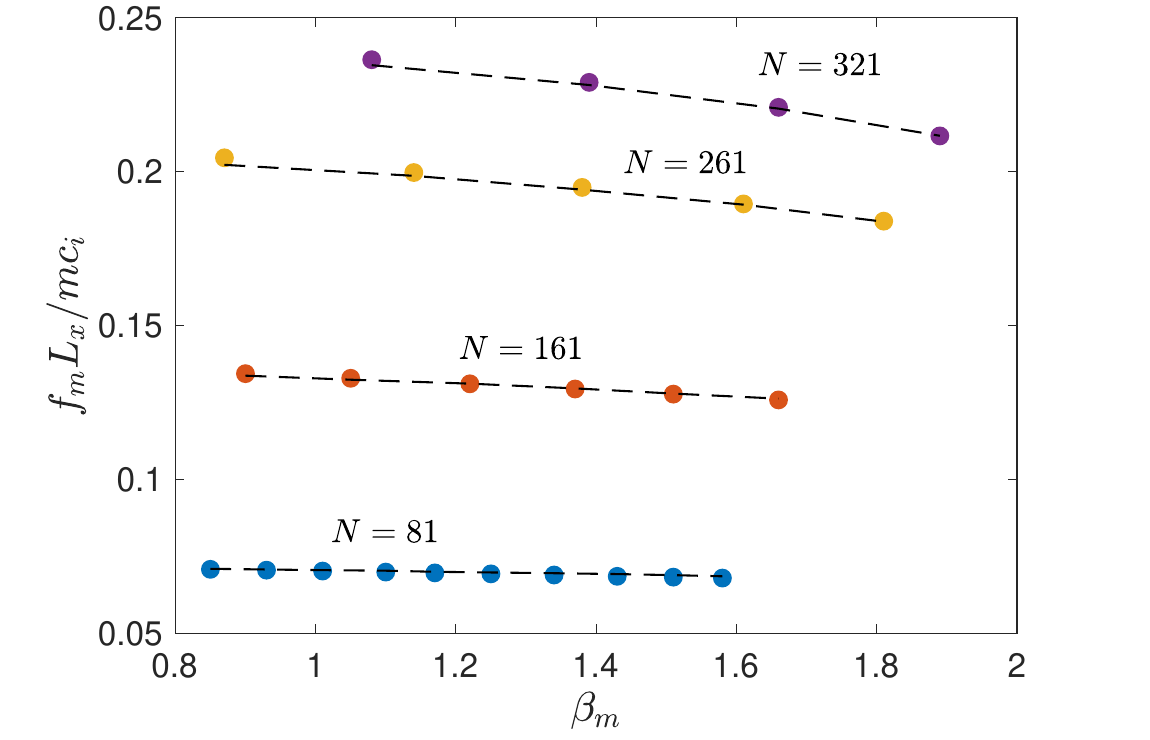}
\caption{ The ratios $f_m L_x/mc_i$, where each data point corresponds to the peaks in $P_N(\beta)$ corresponding to the standing waves with m integer half periods. The dashed lines show $1/2\sqrt{1+\zeta(T)(\pi/N)^2}$. }
\label{lf}
\end{figure}

The EM standing waves shown in Fig. \ref{Bsv} can be described by a subset of eigenmodes in a rectangular 
JJ stack~ \cite{klein,lin-sust,cav1,cav2}:   
\begin{equation}
B(x,z,t)=B_{a}\sin\left(\frac{\pi m x}{L_x}\right)\sin\left(\frac{\pi z}{d}\right)e^{-i\omega_m t},
\label{bu}\\
\end{equation}
 No modes with nodes along $z$ and $y$ were observed in our simulations. The frequencies $f_m=\omega_m/2\pi$ in Eq. (\ref{bu}) for the modes with $m$ half-periods obtained from the Fourier spectra are equidistant in odd $m$. This is shown in Fig. \ref{lf}, where each point corresponding to a peak in $P_N(\beta)$. The so-obtained $\omega_m$ match the eigenfrequencies in the inductively-coupled JJ stack,  $\omega^2(q_x,q_z)=\omega_J^2+c_i^2q_x^2/[1+4\zeta\sin^2(sq_z/2)]$ ~ \cite{klein,lin-sust,sg}, where $c_i=c/\sqrt{\epsilon_c}$,    $q_x=\pi m/L_x$, $q_z=\pi/d$, and $\omega_m\simeq (10-14)\omega_J$ (see Fig. \ref{Bsv} and \ref{Msv}). Because $(\omega_m/\omega_J)^2\gtrsim 10^{2}$ and $sq_z/2\ll 1$, the frequencies $f_m=\omega_m/2\pi$ are equidistant in $m$ and decrease as the thickness of the stack decreases:
\begin{equation}
f_m=\frac{cm}{2L_x\Gamma},\qquad \Gamma=\sqrt{\epsilon_c[1+(\pi\lambda/d)^2]}.
\label{fme}
\end{equation} 
For $L_x=\lambda_c$ and $c_i/\lambda_c=\omega_J$, we get $f_mL_x/mc_i=f_m/m\omega_J=1/2\sqrt{1+\zeta(\pi/N)^2}$.   The ratios $f_mL_x/mc_i$ calculated with the account of the dependence of $\zeta(T)$ on the JJ stack temperature $T(\beta)$ match our numerical data, as shown in Fig. \ref{lf}. Here a weak decrease of $f_mL_x/mc_i$ with $\beta$ results from overheating. 

\begin{figure}[h!]
\includegraphics[width=\columnwidth]{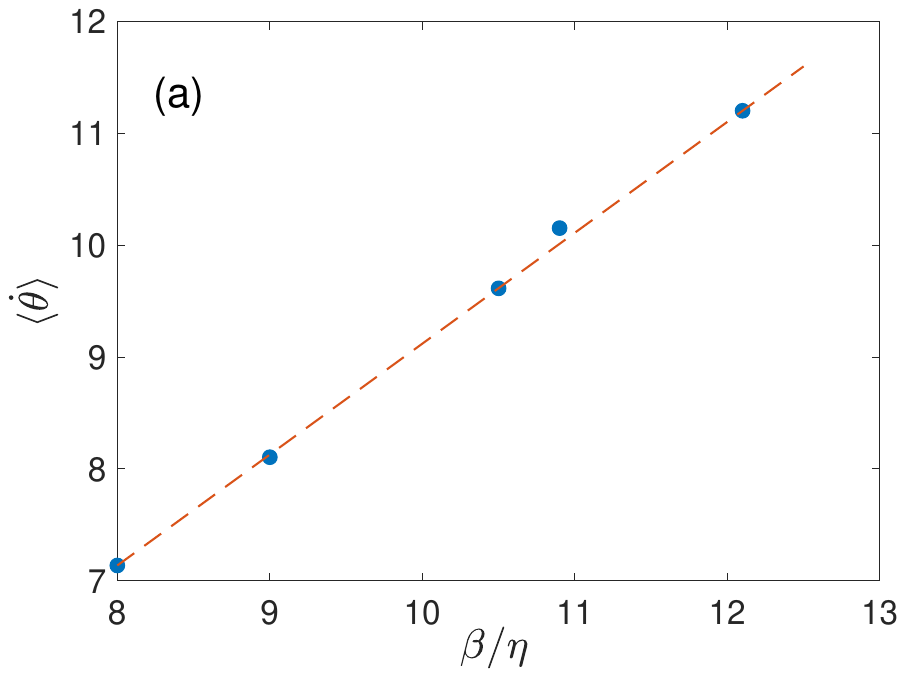}
\includegraphics[width=\columnwidth]{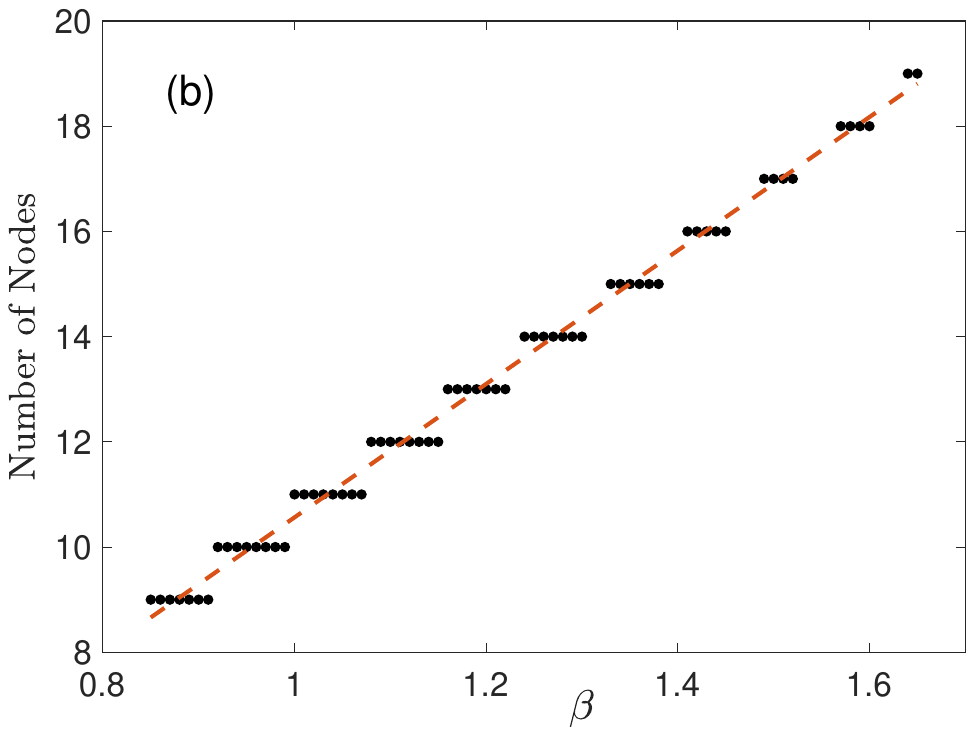}
\caption{(a) A mean time derivative of the phase difference $\langle\dot{\theta}\rangle=\sum_n\int_0^{L_x}\dot{\theta}_n(x,t)dx/NL_x$ in a stack with $N=161$. The dashed line shows 
 $\dot{\theta}=\beta/\eta$ with $\eta=0.1$. (b) The number of nodes in the resonant modes as a function of $\beta$ calculated at $N=161$. The dashed line shows $m=(\beta/\pi\eta)\sqrt{1+(\pi N_c/2N)^2}$ with $N_c=352$ and $\eta=0.1$.}
\label{tdb}
\end{figure}

To get more insight into the resonant frequencies, we plot the averaged $\langle\dot{\theta}\rangle=\sum_n\int_0^{L_x}\dot{\theta}_n(x,t)dx/NL_x$ per JJ at $N=161$ in Fig. \ref{tdb}(a), where  
the dashed line shows $\dot{\theta}_r=\beta/\eta$. The good matching of $\langle\dot{\theta}\rangle$ with $\beta/\eta$ implies that the dc voltage $\hbar\langle\dot{\theta}\rangle/2e$ produced by quasiparticle current causes an oscillating Josephson current $J_c\sin(\beta t/\eta)$ which excites a resonance if $\beta/\eta$ equals one of the mode frequencies $\pi m/\Gamma$. The resulting linear relation $m=\beta\Gamma/\pi\eta$ is consistent with our numerical data shown in Fig. \ref{tdb}(b), where the terraces in $m(\beta)$ are due to the finite widths of peaks and plateaus in $P_N(\beta)$.  The resonance condition $m=\beta\Gamma/\pi\eta$ is equivalent to $2\pi V/\phi_0=\pi cm/\Gamma L_x$, where $V=Js/\sigma_c$ is the dc voltage per JJ. Such resonances occur at applied voltages $V_N(m)=\phi_0cNm/2L_x\Gamma$ or current densities,
\begin{equation}
J_m=\frac{\phi_0 c\sigma_cm}{2 sL_x\Gamma}.
\label{fisk}
\end{equation} 

The magnetic field $B_n(x,t)$ shown in Figs. \ref{Bsv} and \ref{Bmin} vanishes at the edges of the stack. These modes calculated with the boundary conditions $\theta_n'(0)=\theta_n'(L_x)=0$ do not take into account the self-field of transport current. As shown in Appendix \ref{C}, the self-field of a thin rectangular mesa is much smaller than both the mode amplitudes $B_a\sim \phi_0/2\pi s\lambda_c$ and $B_{c1}^\|$. Equations (\ref{bu}) and (\ref{fme}) are derived in a continuum limit in Appendix \ref{D}.

\subsection{Peak radiated power from small stacks}

Now we turn to the dependence of the maximum magneto-dipole power $\tilde{P}_N$ on $N$, where the tilde marks the highest peak in $P_N(\beta)$.  Our calculations for  
$N=21$, $41$ and $81$ show that $\tilde{P}_N$ increases as the power law $\tilde{P}_N\propto N^6$ which gradually slows down at $N>81$: $\tilde{P}_{161}\simeq 37\tilde{P}_{81}$, $\tilde{P}_{261}\simeq 6\tilde{P}_{161}$ and $\tilde{P}_{321}\simeq 1.85 \tilde{P}_{261}$. 

\begin{figure}[h!]
\includegraphics[width=\columnwidth ]{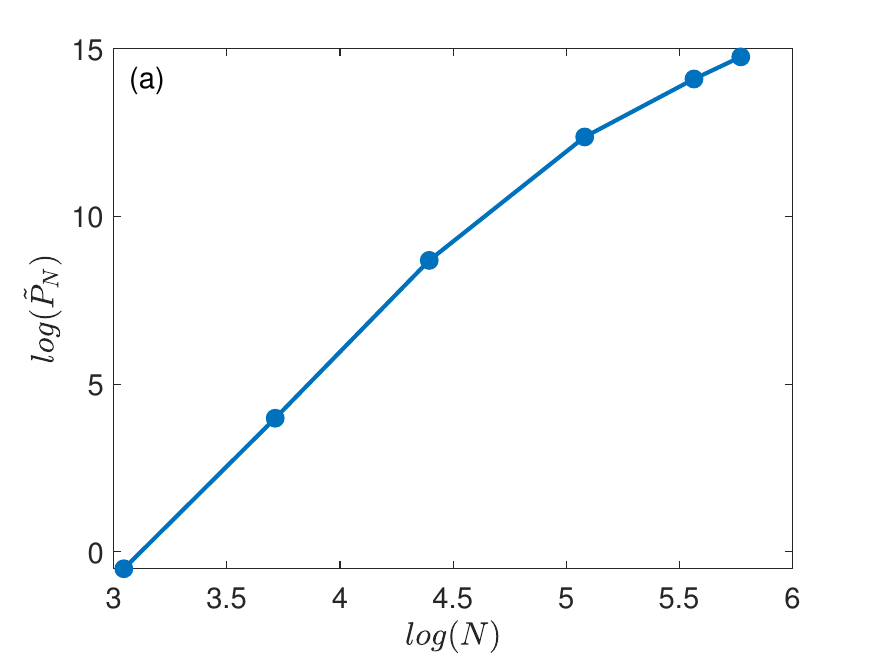}
\includegraphics[width=\columnwidth ]{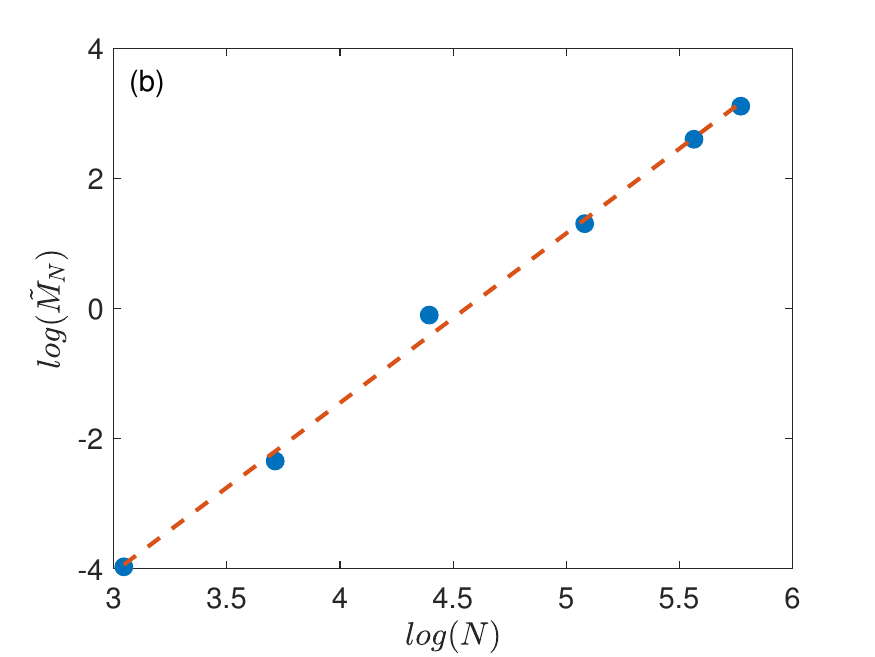}
\caption{The log-log plots of the peak radiation power $\tilde{P}_N$ (a) and the Fourier amplitude $\tilde{M}_N$ (b) calculated for different numbers of JJ layers in the stack. The dashed line in (b) describes the power law $\tilde{M}_N\propto N^{2.58}$. }
\label{logp}
\end{figure}

\begin{figure}[h!]
\includegraphics[width=\columnwidth ]{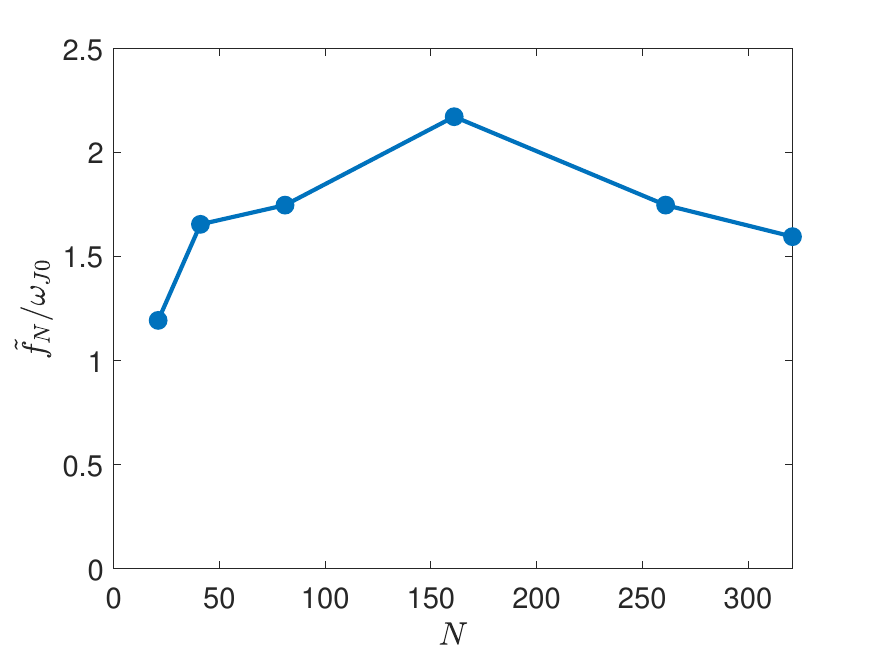}
\caption{The resonant frequency $\tilde{f}_N$ at the highest peaks in the radiated powers as a function of $N$.  }
\label{freq}
\end{figure}

Figure \ref{logp} summarizes the dependencies of $\tilde{P}_N$ and   
$\tilde{M}_N$ on the number of JJs.  One can see that $\tilde{P}_N\propto N^6$ persists up to $N= 81$, while $\tilde{M}_N\propto N^{2.56}$ persists up to $N=321$. Here $\tilde{M}_N$ is proportional to the amplitude of magnetic flux in the odd resonant mode $\Phi=s\sum_n\int B_n(x,t)dx\sim b_m B_0sNL_x/m$, where $m(\beta)$ is the number of nodes and $b_m(\beta)$ is the amplitude of $B_n(t)$ in units of $B_0$. 
The behavior of $\tilde{P}_N$ is determined by the dependencies of $\tilde{M}_N$ and the corresponding resonance frequency $\tilde{f}(N)$ on $N$ shown in Fig. \ref{freq}. Because of the increase of $\tilde{f}(N)$ with $N$ at small $N$, the 
power $\tilde{P}_N\propto \tilde{M}_N^2\tilde{f}(N)^4\propto N^6$ increases faster than $\tilde{M}_N^2$, but slows down at 
$N>81$ as $\tilde{f}(N)$ starts decreasing. Stronger overheating in thicker stacks further slows down the increase of $\tilde{P}_N$ with $N$.  

According to Fig. \ref{Tsv}, the overheating at $N\leq 321$ remains moderate and not strong enough to cause a multivalued $T(\beta)$ necessary for the formation of hotspots ~\cite{gm} in thicker mesas. Yet the decrease of $J_c(T)$ and $\omega_J(T)$ with $T(N)$ diminishes the magnitude and frequency of oscillations of $M(t)$. The interplay of the increase of $M_N$ with $N$ due to increasing magnetic flux in J vortices and decrease of $M_N$ and $\omega_J(N)$ due to increasing $T(N)$ as the stack gets thicker, can 
produce a maximum radiation output at an optimum thickness, although it is not yet reached in Fig. \ref{logp}  
as solving Eqs. (\ref{phase})-(\ref{temp}) with  $N>321$ becomes very time consuming. There are experimental evidences that the power of THz emission from Bi-2212 mesas increases for thicker Bi-2212 crystals ~\cite{thick}. 

\section{One trapped vortex per junction}
\label{mv}

Given that the magnitudes of $M(t)$ at $\beta<\beta_s$ increase with the number of trapped vortices (see Fig. \ref{shuttle}(b)), one may expect that trapping several vortices could produce stronger excitation of the resonant mode and higher $P_N(\beta)$. To investigate this possibility, we solved Eqs. (\ref{phase})-(\ref{temp}) at $\eta=0.1$ with the initial $\theta_n(x,0)$ describing a tilted chain of vortices with one vortex per JJ. Simulations show that, after a small  current $\beta\ll 1$ is applied, this vortex configuration buckles and evolves to a non-periodic structure, consistent with instabilities predicted for vortex structures interacting with a resonant mode in a finite stack \cite{artem,machid,kosh}. We observed that few vortices exited the stack, some accelerated at the edges producing V-AV pairs similar to that is shown in Fig. \ref{fig5}.

As $I$ is increased, trapped vortices driven along different JJs start producing V-AV pairs in the bulk above different threshold currents $\beta_s(n)$ which depend on the positions of JJs in the stack. Here $\beta_s(n)$ is minimum for the central JJ and increases for JJs being closer to the current leads. This happens because the flux $\phi(u_n)$ (see Eq. (\ref{phi})) and the Cherenkov wake in a peripheral vortex is reduced stronger by its AV image than for the vortex in the central JJ. This variation of $\beta_s(n)$ across the stack and multiple degrees of freedom associated with the relative motion of trapped vortices could hinder the synchronization of JJs. Yet our simulations have shown that the resonant mode stimulated by the V-AV pair production by multiple trapped vortices does occur above a threshold current. Such multi-vortex simulations are more time consuming than for a single vortex, so we restricted ourselves to the JJ stacks with $N\leq 161$. 

\begin{figure} [h!]
\includegraphics[width=\columnwidth ]{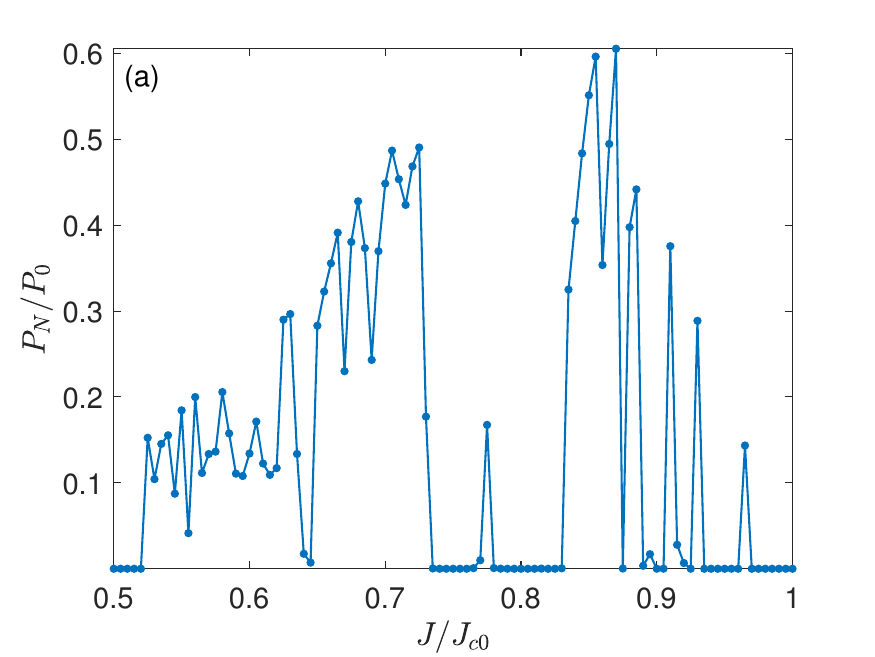}
\includegraphics[width=\columnwidth ]{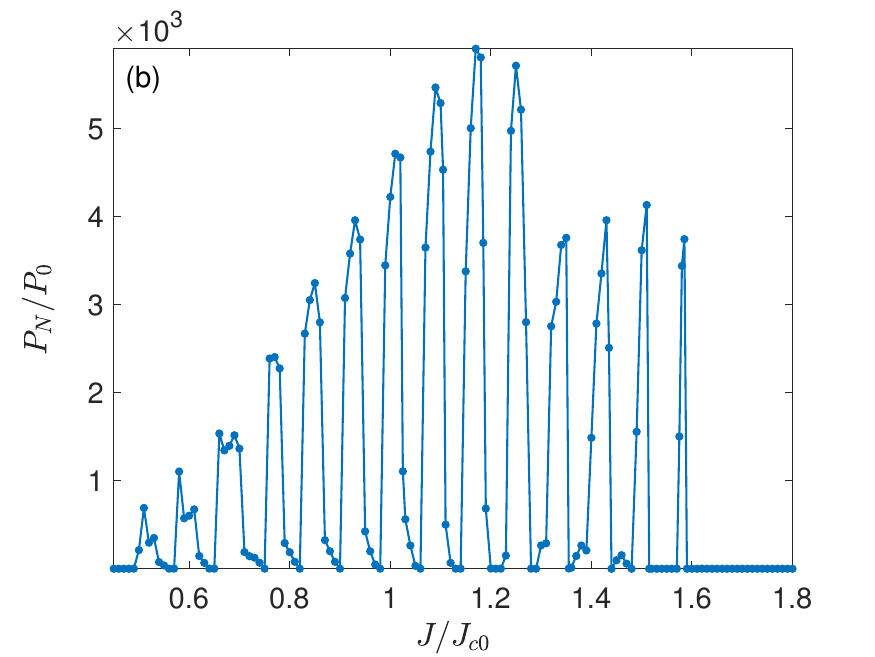}
\includegraphics[width=\columnwidth ]{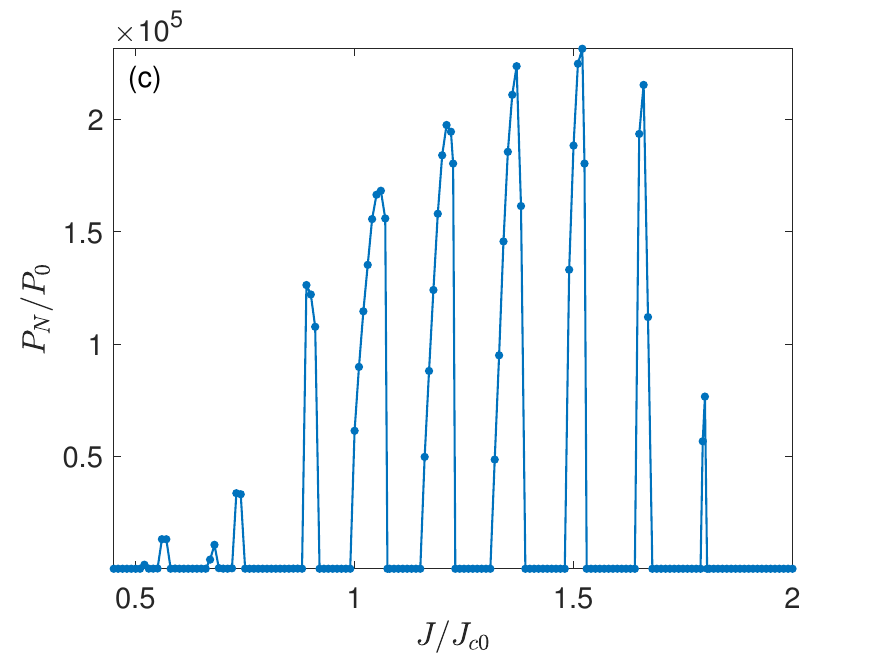}
\caption{The radiated power $P_N(\beta)$ for one trapped vortex per JJ for stacks with: (a) $N=21$, (b) $N=81$, (c) $N=161$.}
\label{Pmv}
\end{figure}

\begin{figure}[h!]
\includegraphics[width=\columnwidth ]{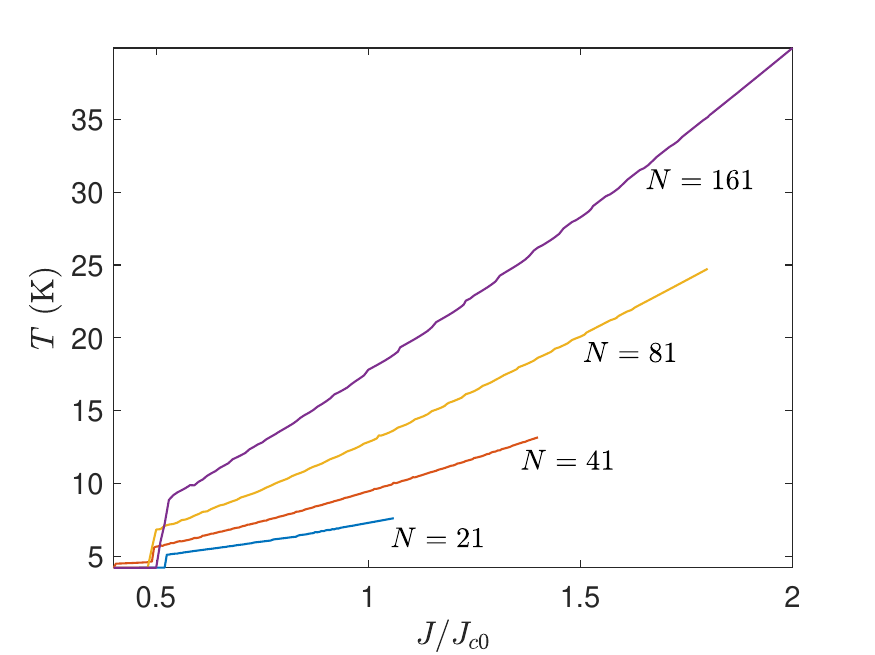}
\caption{$T(J)$ curves for one trapped vortex per JJ for different number of layers. The initial jump is smeared out compared to that in Fig. \ref{Tsv}. }
\label{Tmv}
\end{figure}

\begin{figure}[h!]
\includegraphics[width=\columnwidth ]{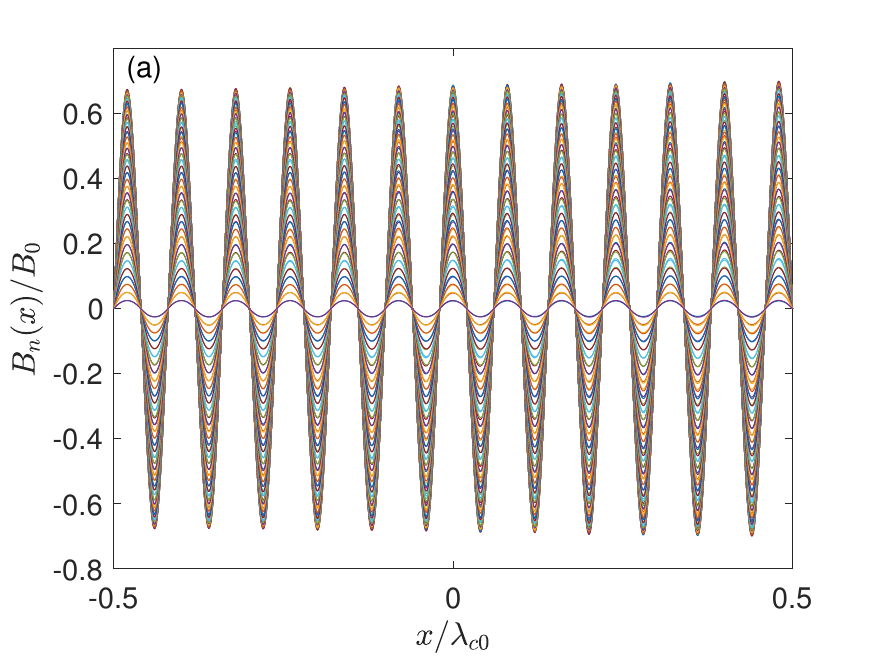}
\includegraphics[width=\columnwidth ]{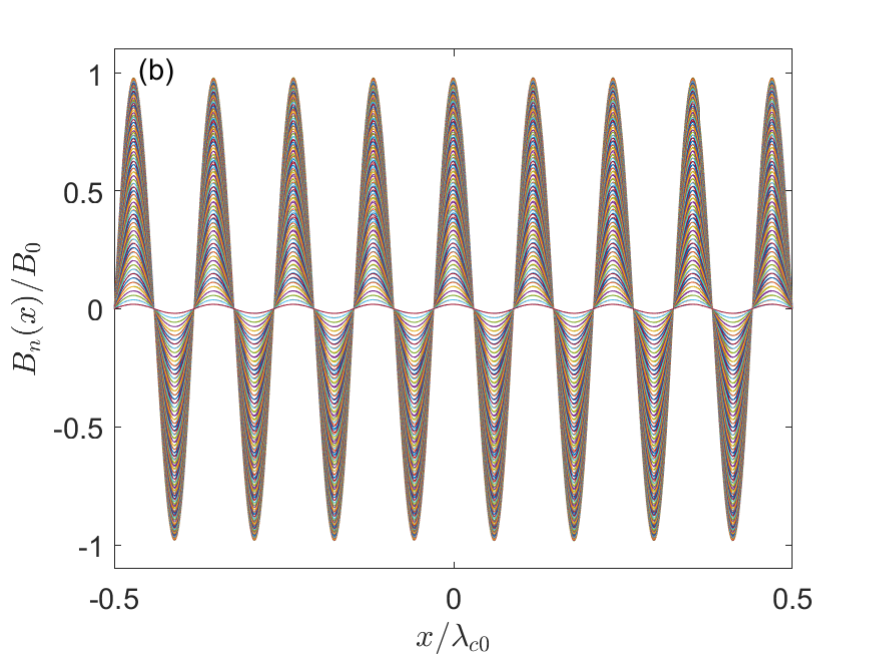}
\caption{Snapshots of resonant modes $B_n(x,t)$ at the maximum radiation power for one trapped vortex per JJ at $N=81$ and $\beta=1.17$ (top) and $N=161$ and $\beta=1.52$ (bottom).}
\label{Bmv}
\end{figure}

\begin{figure}[h!]
\includegraphics[width=\columnwidth ]{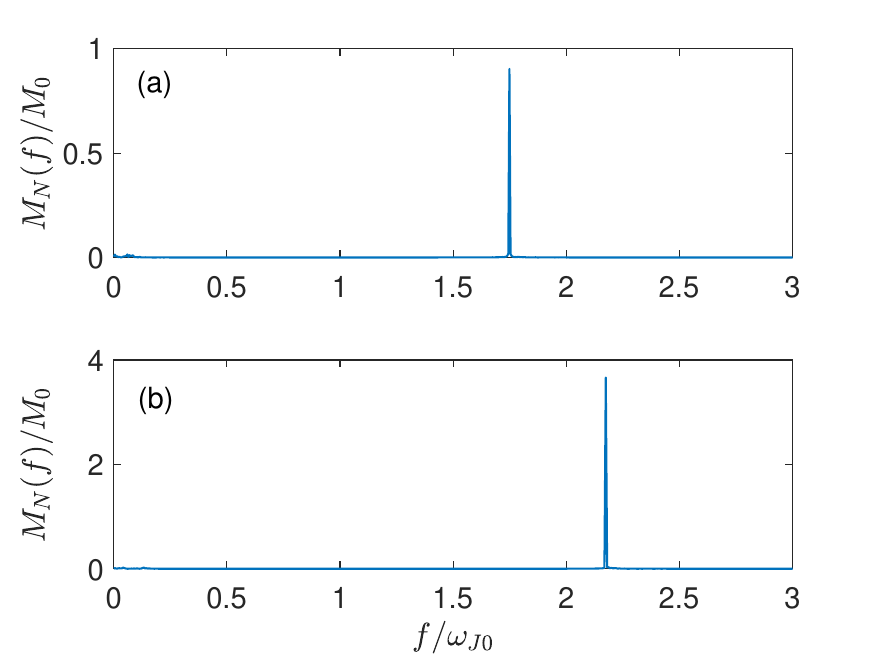}
\caption{Fourier spectra of $M(t)$ at the maximum radiation power for one trapped vortex per JJ and (a) $N=81$ at $\beta=1.17$ and (b) $N=161$ at $\beta=1.52$.}
\label{Mmv}
\end{figure}

\begin{figure}[h!]
\includegraphics[width=\columnwidth ]{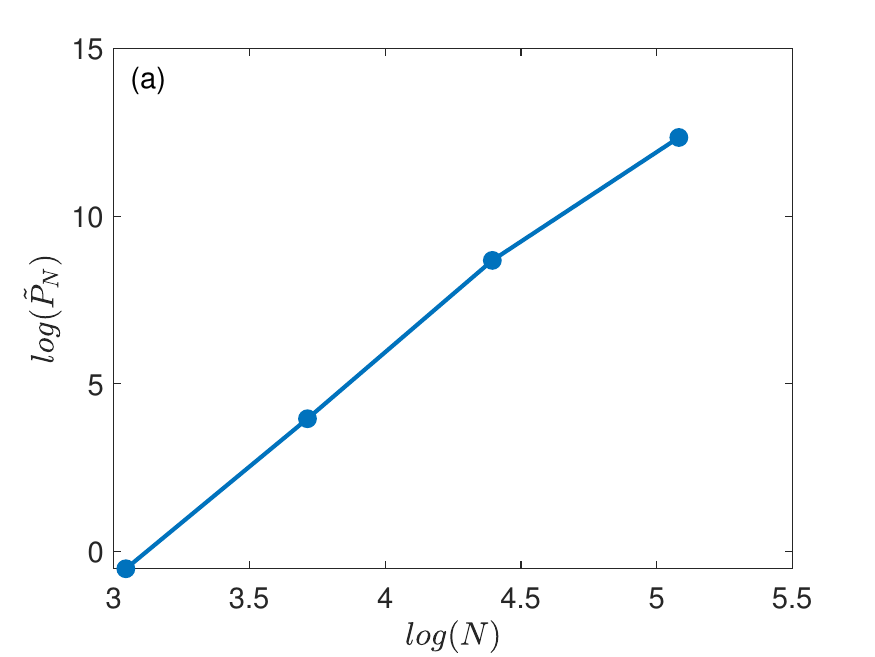}
\includegraphics[width=\columnwidth ]{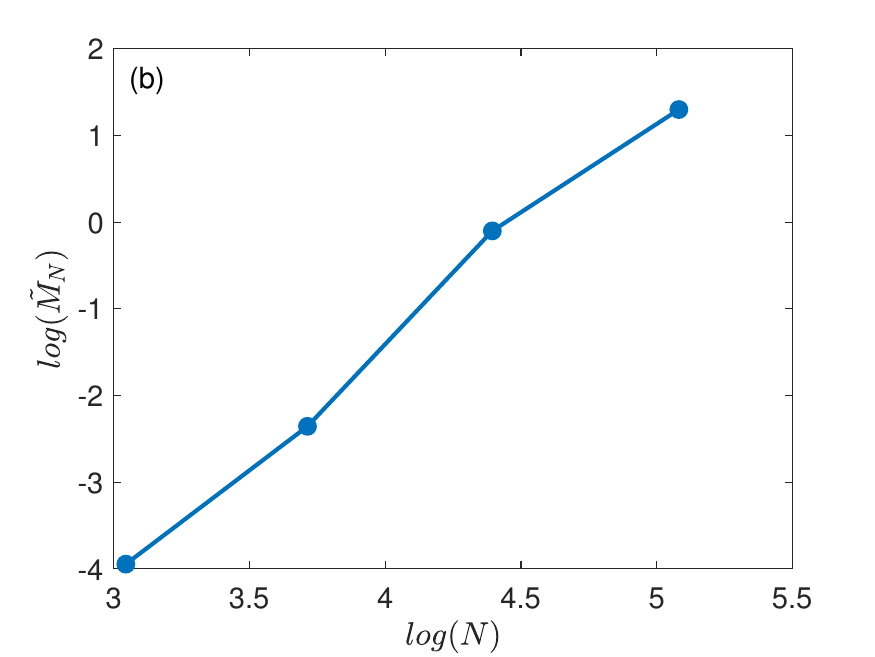}
\caption{The log-log plot of the maximum radiation power $\tilde{P}_N$ (a) and the respective amplitude $\tilde{M}_N$ (b) versus the number of JJs for one trapped vortex per JJ.}
\label{logpmv}
\end{figure}

Figures \ref{Pmv}-\ref{logpmv} summarize our numerical results for one trapped vortex per JJ. The transition to the synchronized JJs is evident from the strong peaks in $P_N(\beta)$, a stepwise increase of the stack temperature $T_N(\beta)$ above a current threshold and the snapshots of $B_n(x,t)$.  The resonant modes $B_n(x,t)$ are stimulated by the multi-vortex shuttles in the stack, the peaks in $P_N(\beta)$ corresponding to the odd number of nodes, as shown in Figs. \ref{Bmv}. The resonant frequencies satisfy Eq. (\ref{fme}) like those shown in Fig.   \ref{lf} for a single vortex.  Yet the number of trapped vortices affects the behavior of $T_N(\beta)$ and $P_N(\beta)$.  For instance, the temperature jump in Fig. \ref{Tmv} is broader than in Fig. \ref{Tsv}, and the distributions of  peaks in $P_N(\beta)$ in Fig. \ref{Pmv} are clearly different from those in Fig. \ref{Psv} for a single vortex. 

Despite the differences in $P_N(\beta)$ for the single and multivortex cases, their maximum $\tilde{P}_N$ corresponding to the highest peaks in $P_N(\beta)$, the respective resonance frequencies and overheating temperatures turned out to be very close, as one can see by comparing Figs. \ref{Msv} and \ref{Mmv} and Figs. \ref{logp} and \ref{logpmv}.  Not only did we observe the same power-law dependencies of $\tilde{P}(N)\propto N^6$ and $\tilde{M}(N)\propto N^{2.58}$ at $N\leq 161$, but also the nearly identical snapshots of $B_n(x,t)$ shown in Figs. \ref{Bsv} and \ref{Bmv}, so that the main characteristics at the maximum radiation output turned out to be independent of the number of trapped vortices.  


\section{Single-mode radiation}

For the parameters $L_x=\lambda_c\simeq 300\,\mu$m, $\epsilon_c=12$, $\omega_{J0}=c/\lambda_{c0}\sqrt{\epsilon_c}\simeq 0.3$ THz used in our simulations, $P_N(\beta)$ is peaked at $f_m\simeq 2\omega_{J0}\simeq 0.6 $ THz (see Fig. \ref{freq}) and $\omega_m\simeq 3.8$ THz. The evaluation of $P_N$ in the dipole approximation captures the qualitative behavior of $P_N(\beta)$ but is applicable for stacks with $L_{x,y}\lesssim c/\omega_m\simeq 80\,\mu$m.  Yet we can use our numerical results for $B_n(x,t)$ valid for any $L_{x,y}$ to evaluate $P_N$  produced by single resonant modes defined by  Eqs. (\ref{bu})-(\ref{fme}) in larger stacks with $L_{x,y}\gtrsim c/\omega_m$. 
To do so, we express $P_N(\beta)$ in terms of the calculated mode amplitude $B_a$ for a rectangular stack by evaluating the far field single-mode radiation vector potential ${\bf A}({\bf R})e^{-i\omega_mt}$ at large distances $R\gg k^{-1}$, where $k=\omega_m/c$ is the radiation wave vector in free space ~ \cite{jack,cav1}:
\begin{equation}
{\bf A}({\bf R})=\frac{\mu_{0}e^{ikR}}{4\pi R}\int_{V}{\bf J}({\bf r})e^{-i{\bf kr}}d^{3}{\bf r}.
\label{ar}
\end{equation}
As shown in  Appendix \ref{D}, the current densities $J_x({\bf r})e^{-i\omega_mt}=-\mu_0^{-1}\partial B/\partial z$ and $J_z({\bf r})e^{-i\omega_mt}=\mu_0^{-1}\partial B/\partial x$ in Eq. (\ref{ar}) can be obtained from Eq. (\ref{bu}), where $J_x$ is determined by in-plane supercurrents and $J_z$ by polarization currents.  In this case the differential radiation power $dP$ within the solid angle $d\Omega$ for modes with $(\omega_m/\omega_J)^2\gg 1$ and $\eta\ll 1$ are given by:
\begin{gather}
\frac{dP}{d\Omega}=\frac{2cB_a^2L_x^2d^2(k_{z}^{2}+k_{x}^{2})k^{2}}{\pi^6\mu_{0}m^2k_{y}^{2}} \times \nonumber \\
\sin^{2}\left(\frac{k_{y}L_{y}}{2}\right)\cos^{2}\left(\frac{k_{x}L_{x}}{2}\right), \quad \mbox{odd m}. 
\label{po}
\end{gather}
\begin{gather}
\frac{dP}{d\Omega}=\frac{2cB_a^2L_x^2d^2(k_{z}^{2}+k_{x}^{2})k^{2}}{\pi^6\mu_{0}m^2k_{y}^{2}} \times 
\nonumber \\
\sin^{2}\left(\frac{k_{y}L_{y}}{2}\right)\sin^{2}\left(\frac{k_{x}L_{x}}{2}\right), 
\quad \mbox{even m}. 
\label{pe}
\end{gather}
For small stacks with $L_{x,y}<c/\omega_m$, Eqs. (\ref{po}) and (\ref{pe}) can be expanded in $k$, giving the dipole radiation power $P_o\propto k^4$ at odd $m$ and quadruple radiation, $P_e\propto k^6\ll P_o$ at even $m$. In spherical coordinates with $k_y=k\cos\chi$, $k_x=k\sin\chi\cos\varphi$ and $k_z=k\sin\chi\sin\varphi$, the angular dependence of $dP_o(\chi,\varphi)$ for odd $m$ becomes:  
\begin{gather}
\frac{dP_o}{d\chi}=\frac{ck^{4}L_{x}^{2}L_y^2d^{2}B_{a}^{2}} 
{\pi^5\mu_{0}m^{2}}\sin^3\chi.
\label{pos}
\end{gather}  
Integration in Eq. (\ref{pos}) gives the total power: 
\begin{gather}
P_o=\frac{4\omega_m^{4}L_{x}^{2}L_y^2d^{2}B_{a}^{2}} 
{3c^3\pi^5\mu_{0}m^{2}}=\frac{\mu_0\langle \ddot{M}^2\rangle}{6\pi c^3},
\label{pot}
\end{gather} 
where $\langle \ddot{M}^2\rangle=M_a^2\omega^4/2$, and $\mu_0M_a = L_y\int_0^{L_x}dx\int_0^d dz B(x,z) = 4B_aL_xL_yd/\mu_0\pi^2m$.  Equations (\ref{pos}) and (\ref{pot}) reproduce $dP/d\Omega$ and $P$ of magneto-dipole radiation used above. Taking   
$\omega_m$ and $\Gamma$ from Eq. (\ref{fme}), we write Eq. (\ref{pot}) in the form:
\begin{equation}
P_o=\frac{cN^2\phi_0^2L_y^2m^2b_a^2}{3\pi^3\mu_0L_x^2\lambda_c^2\epsilon_c^2[1+(\pi N_c/2N)^2]^2}.
\label{pom}
\end{equation}
Here the dimensionless field amplitude $b_a=B_a/B_0$ with $B_0=\phi_0/2\pi s\lambda_{c0}$ and the mode number $m$ for the highest peak in $P$ are to be extracted for each $N$ from the numerical data shown in Fig. (\ref{Bsv}). 

\begin{figure}[h!]
\includegraphics[scale =0.65,trim={150 230 20 100mm},clip]{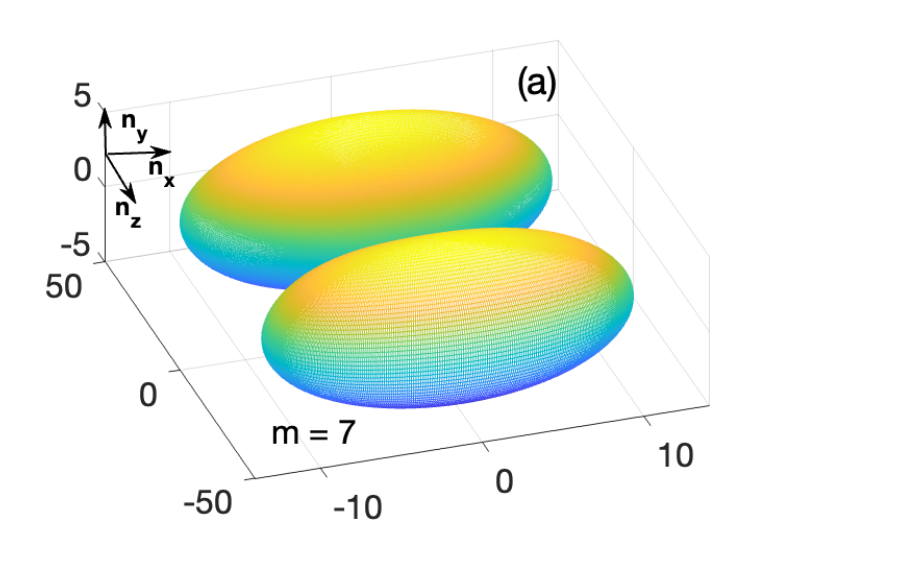}
\includegraphics[scale=0.65,trim={150 260 20 100mm},clip ]{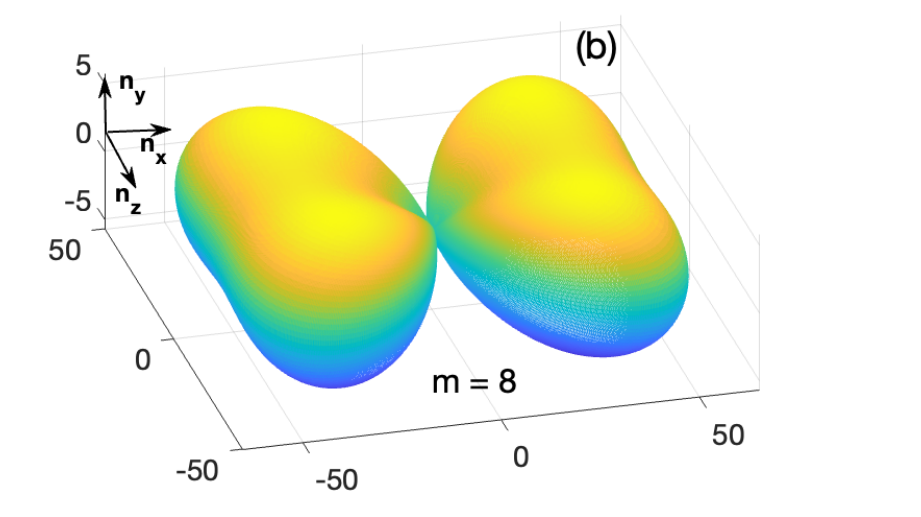}
\caption{Angular distribution of the radiation power $P({\bf n})/P_a$ in the resonant modes with $m=7$ (a) and $m=8$ (b) calculated 
from Eqs. (\ref{po}) and (\ref{pe}) with $L_x=\lambda_c$ and $L_y=4\lambda_c$. Here $P_a=2cB_a^2d^2L_x^2k^2/\pi^6\mu_0m^2$.}
\label{bro}
\end{figure}

For large stacks with $kL_y\gg 1$, not only does the orientational dependence of $dP({\bf n})/d\Omega$ change, but contributions of even and odd modes become of the same order of magnitude. For instance, Fig. \ref{bro} shows markedly different $P_o({\bf n})$ and $P_e({\bf n})$ for odd and even modes calculated at $L_y=4L_x$ and $L_x=\lambda_c$. One can see that $dP({\bf n})/d\Omega$ along the $y$ axis gets much smaller than $dP({\bf n})/d\Omega$ within the $xz$ plane. This happens in the case of large aspect ratios $L_y/L_x$ for which the total radiated power can be calculated using that $\sin^2(k_yL_y/2)/k_y^2$ in Eqs. (\ref{po}) and (\ref{pe}) is peaked at $k_y=0$, that is, $\chi=\pi/2$. Because radiation is mostly confined near the $xz$ plane, we first evaluate the sheet power $P_\varphi=\int_0^\pi P(\chi,\varphi)\sin\chi d\chi$ 
by setting $\cos\chi \to u=(\chi-\pi/2)\ll 1$, $\sin\chi\to 1$, and using $\int_{-\infty}^\infty\sin^2(ukL_y/2)du/u^2=\pi kL_y/2$ to obtain
\begin{gather}
\frac{dP_\varphi}{d\varphi}=\frac{cmL_yd^{2}B_{a}^{2}}{\pi^2\mu_{0}L_x\Gamma^3} 
\cos^{2}\!\left[\frac{\pi m }{2\Gamma}\cos\varphi\right]\!,\quad \mbox{odd\,m}
\label{posh}\\
\frac{dP_\varphi}{d\varphi}=\frac{cmL_yd^{2}B_{a}^{2}}{\pi^2\mu_{0}L_x\Gamma^3} 
\sin^{2}\!\left[\frac{\pi m }{2\Gamma}\cos\varphi\right]\!,\quad\mbox{even\,m} 
\label{pesh}
\end{gather}
Integration of Eqs. (\ref{posh}) and (\ref{pesh}) gives the total power:
\begin{gather}
P_o=\frac{cmL_yd^{2}B_{a}^{2}}{\pi\mu_{0}L_x\Gamma^3}\left[1+J_0\left(\frac{\pi m}{\Gamma}\right)\right],
\quad \mbox{odd\,m}
\label{post} \\
P_e=\frac{cmL_yd^{2}B_{a}^{2}}{\pi\mu_{0}L_x\Gamma^3}\left[1-J_0\left(\frac{\pi m}{\Gamma}\right)\right], \quad \mbox{even\,m}
\label{pest}
\end{gather}
where $J_0(x)$ is a Bessel function. As $\pi m/\Gamma$ increases, $P_o$ and $P_e$ oscillate, approaching the common factor 
$P_\infty=cmL_yd^{2}B_{a}^{2}/\pi\mu_{0}L_x\Gamma^3$, which can be written in the form:
\begin{equation}
P_\infty=\frac{cN^2\phi_0^2L_ymb_a^2}{4\pi^3\mu_0L_x\lambda_{c0}^2\epsilon_c^{3/2}[1+(\pi N_c/2N)^2]^{3/2}}.
\label{pinf}
\end{equation}
At $\pi m/\Gamma \ll 1$ we have $P_o\to 2P_\infty$ and $P_e\to \pi^2m^2P_\infty/4\Gamma^2\ll P_o$, but $P_e$ can exceed $P_o$ if $\pi m/\Gamma>1$. For instance, if $\epsilon_c=12$, $N=321$ and $N_c=352$, we have $\Gamma=6.9$, $P_o(m=7)\approx 0.68P_\infty(m)$ and $P_e(m=8)\approx 1.4P_\infty(m)$. From Eqs. (\ref{pom}) and (\ref{pinf}), it follows that the power radiation scale  $P_\infty$ from large mesas differs from $P_o$ for small mesas by a factor $ \sim (kL_y)^{-1} \simeq L_x\Gamma/m L_y$. 

\begin{figure}[h!]
\includegraphics[scale =0.44,trim={50 200 20 70mm},clip]{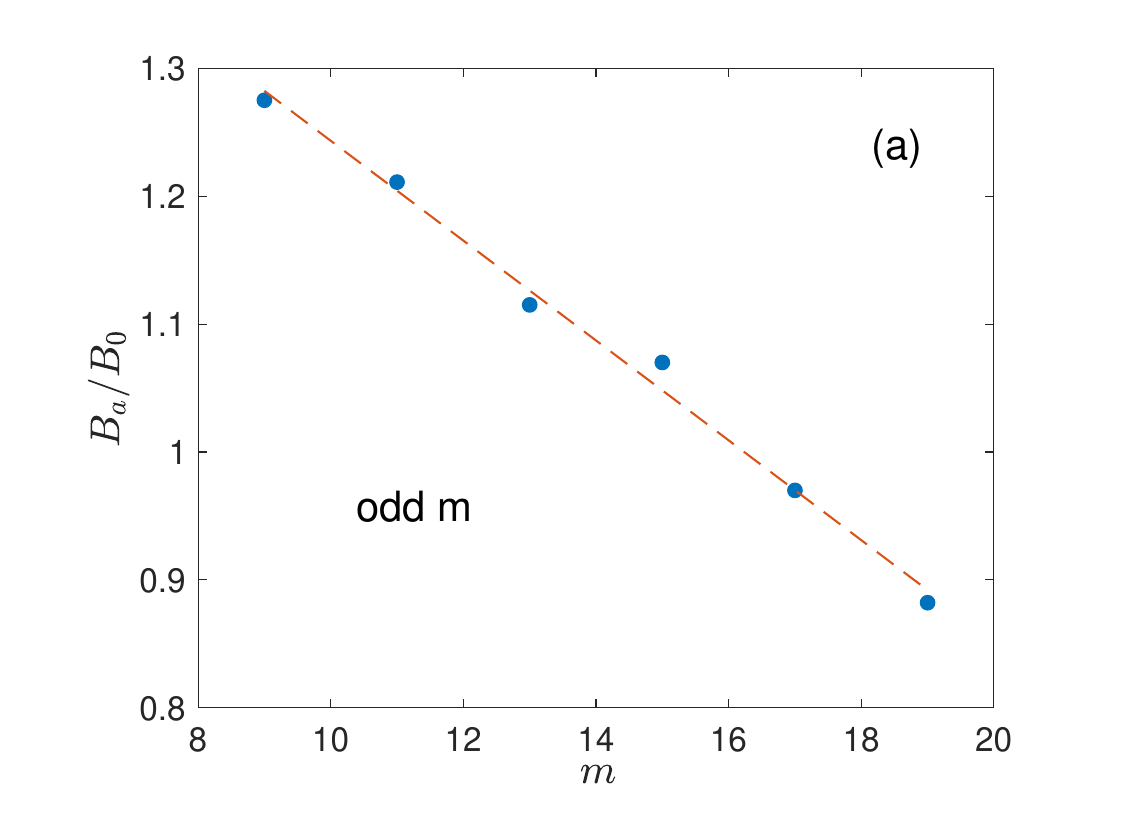}
\includegraphics[scale =0.41,trim={115 220 20 80mm},clip]{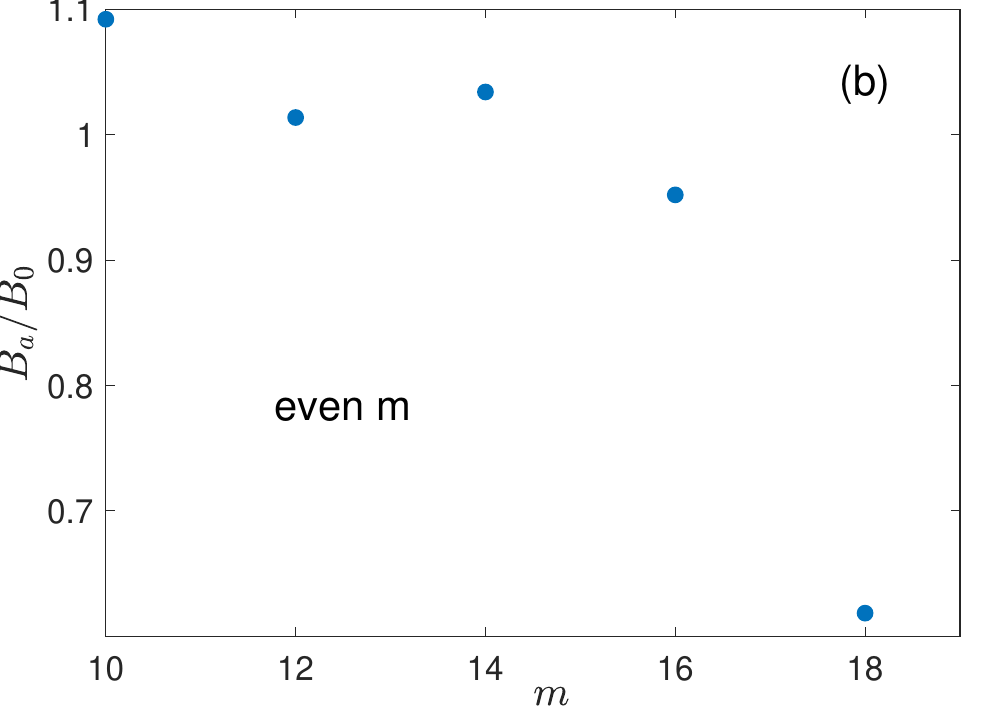}
\caption{Amplitudes of resonant modes as a function of the node number $m$ at $N=161$. Here $B_a(m)$ for odd modes is calculated 
at the radiation peaks, and $B_a(m)$ for even modes is calculated at the middle of the plateaus in $P_N(\beta)$.}
\label{boem}
\end{figure}

\begin{figure}[h!]
\includegraphics[scale =0.45,trim={20 220 20 80mm},clip]{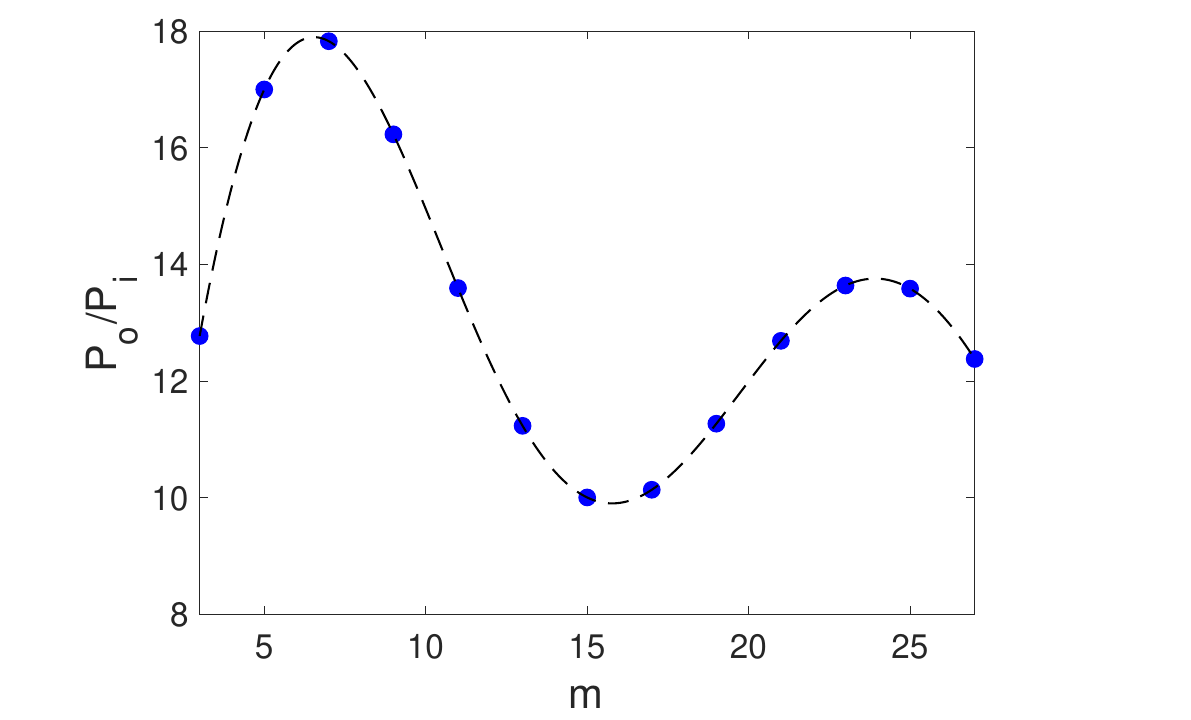}
\caption{Radiated power $P_o(m)$ in units of $P_i= cL_yd^2B_0^2/\pi^2\mu_0^2L_x\Gamma^3$ as a function of odd $m$ calculated from Eq. (\ref{post}) at $N=161$, $\epsilon_c=12$ and $b_a=1.32-0.038(m-8)$ extrapolated from Fig. \ref{boem}(a). } 
\label{radtt}
\end{figure}

To see how $P_o$ and $P_e$  in Eq. (\ref{post}) and (\ref{pest}) depend on the mode numbers $m$, we also need the dependencies of $B_a$ on $m$ to be obtained from the numerical solution of Eqs. (\ref{phase}) and (\ref{Bn}).  Such $B_a(m)$ for even and odd modes at $N=161$ are shown in Fig. \ref{boem}. Here $B_a$ for odd $m$ was calculated at the peaks of $P_N(\beta)$ and $B_a$ for even $m$ was calculated in the middle of the plateaus in $P_N(\beta)$ shown in Fig. \ref{Psv}(b). The evaluation of $B_a$ for even $m$ is ambiguous as $B_a$ varies along the plateau. For odd modes, $b_a\approx 1.32-0.038(m-8)$ decreases linearly with $m$ so  $mb_a^2$ in Eq. (\ref{pinf}) first increases with $m$, reaches  maximum $mb_a^2\approx 16.7$ at $m=14$ and decreases with $m$ at $m>14$. However, interplay of non-monotonic 
$m$-dependencies of $mb_a^2$ and $1+J_0(\pi m/\Gamma)$ produces an oscillatory $P_o(m)$ with maxima at $m=7$, and $m=23$, as shown in Fig. \ref{radtt}. The position and the height of the maxima in $P_o(m)$ depend on the length of the stack, $L_x$.

Both Eq. (\ref{pom}) and (\ref{pinf}) give a gradual change of $P(N)$ from a very rapid increase with $N$ in thin JJ stacks $(N\lesssim N_c$ to a slower increase of $P(N)$ in thicker stacks with $N\gtrsim N_c$. The products $m^2b_a^2$ in Eq. (\ref{pom}) and $mb_a^2$ in Eq. (\ref{pinf}) also depend on $N$ but they vary much slower than the power-laws $P\propto N^6$ at $N\lesssim N_c$ and $P\propto N^2$ at $N\gtrsim N_c$. For small mesas, $m^2b_a^2$ varies from $73$ to $266$ within the range $21<N<321$ in which $P$ increases by $5-6$ orders of magnitude.  In larger mesas Eq. (\ref{pinf}) suggests a gradual transition from $P\propto N^5$ at $N\lesssim N_c$ to $P\propto N^2$ at $N\gtrsim N_c$. Here the product $mb_a^2$ extracted from the numerical data shown in Fig. \ref{Bsv} varies within the range $10-14$ as $N$ increases from $81$ to $321$.  Another source of $N$ dependence of $P_\infty$ comes from overheating which increases $\lambda_c(T)$ and $N_c(T)$. 

\section{Discussion}
\label{sec:disc}
This work shows that the bouncing vortex trapped in a layered superconductor can stimulate the V-AV pair production and resonant modes of JJ stack at a subcritical dc current $I_s<I_c$. The giant amplification of the radiation power at $N\gg 1$ comes from several effects. First, the explosive V-AV pair production caused by Cherenkov instability of the vortex shuttle with low bouncing frequency $f_v\sim (s\lambda_c/\lambda L_x)\omega_J$ excites resonant modes of the JJ stack with frequencies $f_m \sim \omega_J \lambda_c m/L_x$ some 2-3 orders of magnitude higher than $f_v$. This increases the radiation power by the factor $\sim (f_m/f_v)^4\sim 10^8-10^{12}$.  Furthermore, the amplitude of magnetic oscillations increases very rapidly with the number of layers due to better synchronization of JJs caused by the increase of the magnetic flux in $J$ vortices and the number of produced V-AV pairs at $N\lesssim 2\lambda(T)/s\sim 300-500$.   

 Our simulations of up to 321 junctions have shown that resonant modes stimulated by the V-AV shuttle produce peaks in the radiation power $P_N(J)$ at $J>J_s$, yet increasing the number of trapped vortices affects weakly the maximum radiation output and the overheating.  To estimate the magnitude of $P_N(m)$ for the mode with $m=7$ in the JJ stack with $N=321$ and $\eta=0.1$, we take $b_a\approx 1.2$ from Fig. \ref{Bsv}(c),  $L_y/L_x=4$, $L_x=\lambda_c=300\, \mu$m, $\epsilon_c=12$. In this case $P_\infty\simeq 0.4\, \mu$W. If the factor $mb_a^2$ remains weakly-dependent on $N$ at $N>321$, Eqs. (\ref{post})-(\ref{pinf}) suggest  $P\sim 10\,\mu$W in a $2-3\,\mu$m thick Bi-2212 mesa. Calculation of the actual radiation output may require taking into account a more complicated boundary condition ~\cite{bc1,bc2,bc3,bc4} at the edges of the stack instead of $\theta_n'(0)=\theta'(L_x)=0$ used here. Given the large impedance mismatch between the JJ stack and the vacuum captured by our boundary conditions, the more consistent boundary condition taking into account matching of incoming and radiated EM waves at the edges may not qualitatively change our results but would require numerical calculations including coupling of the mesa with the surrounding structures ~\cite{thz1,thz2,thz3,thz4,thz5,machida,tachiki,lin-sust}.  
 
The power $P_N$ depends essentially on the magnitude and the temperature dependence of the damping parameter $\eta$ which is affected by doping ~\cite{sigc1,sigc2,sigc3}. Decreasing $\eta$ by doping would reduce both the onset $J_s(\eta)$ of V-AV pair production (see Fig. \ref{Jsv}) and the currents $J_m$ of intrinsic resonances given by Eq. (\ref{fisk}). At $\eta\ll 1$ the trapped vortex can excite high-frequency modes with $m \gg 1$ at $J=J_s$, thus increasing $P_N$ as compared to $P_N$ calculated here for $\eta=0.1$  and stimulating THz emission at subcritical currents at which overheating is reduced. Doping also affects a dissipative contribution of in-plane quasiparticle currents~ \cite{artem,thz5} neglected in Eqs. (\ref{phase})-(\ref{Bn}).  The in-plane  damping is most pronounced in optimally-doped Bi-2212 ~\cite{thz5,coupl1} in which it can increase $J_s$ and mitigate the V-AV pair production. We consider here the most favorable for the V-AV pair production case of weak inter plane damping and negligible in-plane damping which may pertain to underdoped Bi-2212.

Utilizing trapped J vortices to stimulate THz emission offers an opportunity to optimize the radiation output by tuning the field cooling of the mesa through $T_c$ in a weak magnetic field.  For a thick Bi-2212 crystal with $d>2\lambda$, the parallel lower critical field $B_{c1}^\|= (\phi_0/4\pi\lambda\lambda_c)[\ln(\lambda/s)+1.12]\simeq B_{c1}^\|(0)(1-T^2/T_c^2)$~ \cite{clem} not only vanishes at $T_c$ but $B_{c1}^\|(0)\simeq 0.12$ Oe at $T\ll T_c$ is below the Earth magnetic field. Thus, trapping an optimum number of vortices upon cooling the mesa through $T_c$ requires appropriate screening and alignment of the Earth field relative to the ab planes. However, $B_{c1}^\|(T)$ at $T\to T_c$ exhibits a dimensional crossover from the bulk limit at $d>2\lambda(T)$ to a thin film limit at $d<2\lambda(T)$ in which a weakly $T$ dependent  
$B_{c1}^\|\simeq (2\phi_0\lambda/\pi \lambda_cd^2)\ln(d/s)$ can significantly exceed the bulk $B_{c1}^\|$ because of the reduction of magnetic flux of the vortex ~\cite{vf0,vf1,vf2,vf3} (see Appendix \ref{C}). In thin JJ stacks with $d<2\lambda(0)$, the lower critical field is much higher than the bulk $B_{c1}^\|(T)$ at all $T$.  

THz emission stimulated by trapped Josephson vortices can be affected not only by the magnitude but also by the orientation of the ambient field ${\bf H}$ relative to the ab planes of Bi-2212 crystals. Good  alignment of ${\bf H}$ with the ab planes may be essential to mitigate detrimental penetration of pancake vortices which could deteriorate synchronization of the intrinsic JJs.

\section*{Acknowledgments}
This work was supported by DOE under Grant No. DE-SC0010081-020 and by the AFOSR 
under grant No. FA9550-17-1-0196.


\appendix

\section{Matrix form of Eq.(\ref{phase}) and the boundary conditions.} \label{A}

The current density across the $n$-th Josephson junction being between the $(n-1)$-th and $n$-th superconducting layers is given by
\begin{equation}
J_n^z = \alpha_J(T)\sin\theta_n +\eta \dot{\theta_n}+\ddot{\theta_n}
\label{A1}
\end{equation}
In turn, the Maxwell equations give~\cite{bul,kleiner,machida,tachiki}
\begin{equation}
\theta_n^{\prime\prime}=J_n^z-\zeta(T)(J_{n+1}^z-2J_n^z+J_{n-1}^z)
\label{A2}
\end{equation}
Combining Eqs. (\ref{A1}) and (\ref{A2}) result in Eq. (\ref{phase}).

For the mesa geometry shown in Fig. \ref{mesa}, $J_{n-1}^z$ and $J_{n+1}^z$ for the top and the bottom junctions respectively, is  replaced with the injected dc current density $J$. Then Eqs. (\ref{A1}) and (\ref{A2}) reduce to the matrix form:
\begin{equation}
\mathbf{\Theta^{\prime\prime}-V=A\cdot(J^z-V)},
\label{A3}
\end{equation}
where $\mathbf{\Theta}=(\theta_1,\theta_2,...,\theta_N)$, $\mathbf{V}=\beta(1,1,...,1)$, $\mathbf{J^z}=(J^z_1,J^z_2,...,J^z_N)$. The matrix $\mathbf{A}$  is given by $A_{i,i}=1+2\zeta(T)$, $A_{i,i+1}=A_{i,i-1}=-\zeta(T)$ and $A_{i,j}=0$ otherwise. The resulting equations were solved by the method of lines~\cite{mdln} which turns the partial differential  Eqs. (\ref{A3}) into  a set of ordinary differential equations in time.

\section{Heat transfer across the mesa} \label{B}

For a thin mesa situated at $0<z<d\ll w$ on a slab of thickness $w$ along $z$ 
and length $L_x$ along $x$, the thermal diffusion equation is
\begin{gather}
C\partial_t T=\partial_x(\kappa\partial_{x}T)+\partial_z\left(\kappa_{c}\partial_{z}T\right)+Q(T)\delta(z),
\label{B1} \\
\partial_{x}T\big|_{x=0}=\partial_{x}T\big|_{x=L_x}=0,
\label{B2}
\end{gather}
where $C(T)$ is the specific heat, $\kappa_{\alpha\beta}(T)$ is a thermal conductivity tensor with the principal values  
$\kappa$ and $\kappa_c$ in a uniaxial crystal with $c\| z$, and $Q(T)$ is 
the power dissipated in the mesa with thermally-insulated sides. 

A mean temperature $T_m$ along the mesa satisfies a stationary equation, $\partial_z(\kappa_c\partial_z T)+Q\delta(z)=0$. 
Integrating this equation gives a constant heat flux $q=-\kappa_c\partial _zT$ at $z>0$. Hence, $q=\int_0^w\kappa_c\partial_z Tdz/w=w^{-1}\int_{T_i}^{T_m}\kappa_cdT$, 
and the condition $Q(T_m)=q$ yields the equation for $T_m$:
\begin{equation}
\frac{1}{w}\int_{T_i}^{T_m}\kappa_c(T)dT=Q(T_m).
\label{B3}
\end{equation}
The temperature $T_i$ at the bottom of the base is determined by 
the boundary condition:
\begin{equation}
\kappa_c(T_i)\partial_zT\big|_{z=w}=Y(T_i^n-T_0^n),
\label{B4}
\end{equation}
Here $Y$ is inversely proportional to the Kapitza contact thermal resistance $R_K=(nYT_0^{n-1})^{-1}$ at $T_i-T_0\ll T_0$, resulting in a temperature jump $T_i-T_0$ between the base and the sample holder maintained at the ambient temperature $T_0$. The exponent $n$ can vary between $3$ and $5$, depending on the interface properties ~\cite{kapitza}.  The equation for $T_i$ readily follows from Eqs. (\ref{B3})-(\ref{B4}):
\begin{equation}
\frac{1}{w}\int_{T_i}^{T_m}\kappa_c(T)dT=Y(T_i^n-T_0^n)
\label{B5}
\end{equation}
  
We evaluate the effect of the Kapitza resistance on $T_m$ at $T_i-T_0\ll T_0$ neglecting the temperature dependence of $\kappa_c(T)$. Then Eqs. (\ref{B3}) and (\ref{B5}) reduce to: 
\begin{gather}
Q(T_m)=(T_m-T_i)\kappa_c/w, 
\label{B6} \\
(T_m-T_i)\kappa_c/w=(T_i-T_0)h_K,
\label{B7}
\end{gather}
where $h_K=nYT_0^{n-1}$. Eqs. (\ref{B6}) and (\ref{B7}) yield:
\begin{gather}
Q(T_m)=\frac{(T_m-T_0)\kappa_c}{w+\kappa_c/h_K}.
\label{B9}
\end{gather}
The Kapitza resistance enhances overheating if $wh_K< \kappa_c$.  The magnitude of $h_K$ depends on many poorly understood factors including  
the effect of the Bi-2212 layered structure on the acoustic mismatch of the base and the substrate and 
details of atomic structure of the interface \cite{kapitza}. Typically $h_K\sim (1-5)\times 10^4$ W/m$^2$K between a metal and the 
liquid He at 4.2 K, in which case $wh_K/\kappa_c\sim 1$ for $w=30\,\mu$m and $\kappa_c=0.6$ W/mK.
Yet because $h_K\propto T^3$ increases faster with $T$ than $\kappa_c\propto T^{0.67}$, the mesa temperature $T=20-40$ K shown in Fig. \ref{Tsv} corresponds to the case $h_K\gg \kappa_c/w$ in which the Kapitza resistance can be disregarded, and Eq. (\ref{B3}) becomes 
\begin{equation}
Q(T_m)=\frac{\kappa_0T_0}{w(a+1)}\left[\left(\frac{T_m}{T_0}\right)^{a+1}-1\right].
\label{B10}
\end{equation} 
This equation with $T_m\to T$ was used in our calculations.

\section{Self-field effects} \label{C}
The in-plane self-field $H_{y}$ of a uniform current density $J$ flowing along $z$ is given by the Biot-Savart law:
\begin{equation}
\!\!H_{y}({\bf r})=J\!\int_V\frac{(x'-x)d^{3}{\bf r}'}{|(x-x')^{2}+(y-y')^{2}+(z-z')^{2}|^{3/2}}
\label{C1}
\end{equation}
We evaluate $H_y$ in the middle of the stack left face parallel to the trapped vortex $(x=0, \,y=L_{y}/2,\,z=d/2)$ by first integrating Eq. (\ref{C1}) with respect to $z'$: 
\begin{gather}
\!\!H_{y}=Jd\!\int_{0}^{L_{x}}\!\!dx'\!\int_{-L_{y}/2}^{L_{y}/2}\!
\frac{x'dy'}{(x'^{2}+y'^{2})\sqrt{x'^{2}+y'^{2}+(d/2)^2}}.
\label{C2}
\nonumber
\end{gather}
The $y'$ integral is then evaluated at $L_{y}\gg L_{x}$: 
\begin{gather}
\!\!\int_{-\infty}^{\infty}\!\frac{dy'}{(x'^2+y'^{2})\sqrt{(d/2)^{2}+x'^2+y'^{2}}}=
\frac{4}{dx'}\tan^{-1}\frac{d}{2x'}.
\label{C3}
\nonumber
\end{gather}
Hence, we obtain $H_y$ at $d\ll L_x$:
\begin{gather}
\!\!\! H_{y}=4J\!\!\int_{0}^{L_{x}}\!\!dx'\tan^{-1}\frac{d}{2x'}\simeq
2Jd\left[\ln\frac{2L_{x}}{d}+1\right].
\label{C4}
\end{gather}
This formula defines the scale of the in-plane self field $B_i\simeq \mu_0H_y(0)$ which is to be compared to the field amplitude $B_a\sim B_0=\phi_0/2\pi s\lambda_c$ in the resonant modes 
shown in Fig. \ref{Bsv}. Using $\mu_0J_c=B_0/\lambda_c$, we recast $B_i$ to
\begin{equation}
B_{i}\simeq 2\beta B_0\frac{d}{\lambda_c}\left(\ln\frac{2L_{x}}{d}+1\right),
\label{C5}
\end{equation}
where $\beta=J/J_c$.  The condition $B_{i}\ll B_0$ requires:
\begin{equation}
\beta\ll \frac{\lambda_c}{2d[\ln(2L_x/d)+1]}.
\label{C6}
\end{equation}
Eq. (\ref{C6}) is satisfied at $L_x=\lambda_c\gg d$ and $\beta\sim 1$. 

Compare now the self-field with the in-plane 
lower critical field  $B_{c1}^\|= (\phi_0/4\pi\lambda\lambda_c)[\ln(\lambda/s)+1.12]$ at $d>\lambda$ ~\cite{clem}.
The condition $B_i<B_{c1}^\|$ requires
\begin{equation}
\beta <\frac{\lambda_c[\ln(\lambda/s)+1.12]}{4N\lambda[\ln(2L_x/d)+1]},\qquad d>2\lambda
\label{C7}
\end{equation}
If $N>N_c(T)=2\lambda/s\simeq 300-500$, Eq. (\ref{C7}) may not be satisfied at $\beta>\beta_{s}$ 
so vortices could penetrate as $\beta$ is increased. This may not significantly affect $P_N$ given the  
weak sensitivity of $P_N$ to the number of trapped vortices. 

If $d<2\lambda$ the condition $B_i<B_{c1}^\|$ is satisfied more easily due to larger  
$B_{c1}^\|\simeq (2\phi_0\lambda/\pi \lambda_cd^2)\ln(d/s)$ in thin films ~\cite{vf0,vf1,vf2,vf3}, where 
the factor $\lambda/\lambda_c\ll 1$  accounts for the anisotropy of  
screening. In this case  $B_i<B_{c1}^\|$ if
\begin{equation}
\beta\lesssim 
\frac{2\lambda\lambda_c\ln(d/s)}{Nd^2[\ln(2L_x/d)+1]},\qquad d<2\lambda.
\label{C8}
\end{equation}
This condition is satisfied if $d=\lambda/2$, $L_x=\lambda_c$, $N=200$, $\lambda_c/\lambda\sim 10^{3}$ and $\beta\simeq 1$, so the 
number of trapped vortices remains constant, as it was assumed in our simulations. 

\section{Single mode radiation} \label{D}
To calculate the far field radiation vector potential in Eq. (\ref{ar}), we express the current density in the mesa ${\bf J}({\bf r})$ in terms of the resonant field  $B$ along $y$. Because $\theta_n(x,t)$ varies slowly over the JJ spacing, we replace the discrete $\theta_n$ and $B_n$ with  smooth functions $\theta(x,z,t)$ and $B(x,z,t)$ satisfying Eqs. (\ref{phase})-(\ref{Bn}) in a continuum limit:
\begin{gather}
\!\!\frac{\partial^2\theta}{\partial x^2}=\bigl[1-\lambda^2\frac{\partial^2}{\partial z^2}\bigr]\bigl[\frac{\alpha}{\lambda_c^2}\sin\theta+\mu_0\sigma_c\frac{\partial\theta}{\partial t}
+\frac{\epsilon_c}{c^2}\frac{\partial^2\theta}{\partial t^2}\bigr],
\label{D1} \\
B-\lambda^2\frac{\partial^2B}{\partial z^2}=\frac{\phi_0}{2\pi s}\frac{\partial \theta}{\partial x}.
\label{D2}
\end{gather}
Here the driving term $\sin\theta$ oscillating with the frequency $\beta\omega_J/\eta$ sets the amplitude of the 
resonant mode calculated numerically. The Josephson and ohmic terms in Eq. (\ref{D1}) 
affect weakly the resonant frequencies $\omega_m\gg \omega_J$ which can be obtained by setting $\alpha=\sigma_c=0$.  In this case 
the eigenfrequencies and eigenfunctions of Eqs. (\ref{D1})-(\ref{D2}) satisfying the boundary conditions 
$\partial_xB(0,z)=\partial_xB(L_x,z)=0$ and $\theta(x,0)=\theta(x,d)=0$ are:
\begin{gather}
\omega_m=\frac{\pi m c}{L_x\Gamma},\qquad 
\Gamma=\sqrt{\epsilon_c[1+(\pi\lambda/d)^2]},
\label{D3} \\
B=B_a\sin\left(\frac{\pi mx}{L_x}\right)\sin\left(\frac{\pi z}{d}\right)e^{-i\omega_m t}.
\label{D4}\\
\theta=-\frac{2B_asL_x\Gamma^2}{m\phi_0\epsilon_c}\cos\left(\frac{\pi mx}{L_x}\right)\sin\left(\frac{\pi z}{d}\right)e^{-i\omega_m t},
\label{D5}
\end{gather}
The in-plane current density $J_x$ is dominated by the Meissner current,   
the displacement current is negligible $(\omega_m\lambda/c)^2=(\pi m\lambda/\Gamma L_x)^2\ll 1$.  In turn,  
$J_z$ is dominated by the polarization current $J_z\simeq (\epsilon_c\hbar/2esc^2\mu_0)\omega_m^2\theta\sim (\omega_m/\omega_J)^2J_c\gg J_c$.  
Using here $\theta$ from Eqs. (\ref{D3})-(\ref{D5}), we obtain  $J_z=\partial B/\partial x$, $J_x=-\partial B/\partial z$, $J_y=0$, where
\begin{gather}
J_x=-\frac{\pi B_{a}}{\mu_{0}d}\sin\left(\frac{\pi mx}{L_x}\right)\cos\left(\frac{\pi z}{d}\right)e^{-i\omega_m t},
\label{D6} \\
J_z=\frac{\pi m B_a}{\mu_0L_x}
\cos\left(\frac{\pi mx}{L_x}\right)\sin\left(\frac{\pi z}{d}\right)e^{-i\omega_m t}.
\label{D7}
\end{gather}
In this mode $\nabla\cdot {\bf J}=0$ so it produces no macroscopic charge densities in the bulk and the surface. 

The single-mode radiation vector potential ${\bf A}({\bf k},R,t)$ is readily obtained from 
Eqs. (\ref{ar}), (\ref{D9}) and (\ref{D10}): 
\begin{gather}
A_{x}=\frac{i\pi B_amk_{z}L_xd[(-1)^me^{-ik_{x}L_{x}}-1]}{2R(k_{x}^{2}L_{x}^{2}-\pi^{2}m^{2})(\pi^{2}-d^{2}k_{z}^{2})k_{y}}\times 
\nonumber \\
\left[1+e^{-idk_{z}}\right]\sin\left(\frac{k_{y}L_{y}}{2}\right)e^{-ik_{y}L_{y}/2+ikR-i\omega_m t},
\label{D8}\\
A_{z}=-A_yk_y/k_z,\qquad A_y=0,
\label{D9}
\end{gather}
where $k=\omega_m/c=\pi m /L_x\Gamma$. Because $dk_z \lesssim d\omega_{m}/c= \pi md/\Gamma L_x\ll1$, we set $e^{-ik_zd}\to 1$. 

The differential radiation power $dP=S({\bf n})R^2d\Omega$ within the solid angle $d\Omega$ is determined by the Poynting vector 
${\bf S}=c|{\bf B}|^2{\bf n}/2\mu_0$ for the far field  
${\bf B}=i{\bf k}\times{\bf A}=i\hat{x}k_yA_z+i\hat{y}(k_{z}A_x-k_xA_z)-i\hat{z}k_{y}A_{x}$ in a plane EM wave propagating along the unit vector ${\bf n}={\bf k}/k$ ~ \cite{jack}:
\begin{equation}
\frac{dP}{d\Omega}=\frac{cR^2}{2\mu_0}[k_y^2(|A_x|^2+|A_z|^2)+|k_zA_x-k_xA_z |^2].
\label{D10}
\end{equation}
Eqs. (\ref{D8})-(\ref{D10}) give:
\begin{gather}
\frac{dP}{d\Omega}=\frac{2cm^2L_{x}^{2}d^{2}B_a^2(k_z^2+k_x^2)k^2}{\pi^2\mu_0(k_{x}^{2}L_{x}^{2}-\pi^{2}m^{2})^{2}k_{y}^{2}}\times
\nonumber \\
\sin^{2}\left(\frac{k_{y}L_{y}}{2}\right)\cos^{2}\left(\frac{k_{x}L_{x}}{2}\right),\quad \mbox{odd\, m}.
\label{D11} \\
\frac{dP}{d\Omega}=\frac{2cm^2L_{x}^{2}d^{2}B_a^2(k_z^2+k_x^2)k^2}{\pi^2\mu_0(k_{x}^{2}L_{x}^{2}-\pi^{2}m^{2})^{2}k_{y}^{2}}\times
\nonumber \\
\sin^{2}\left(\frac{k_{y}L_{y}}{2}\right)\sin^{2}\left(\frac{k_{x}L_{x}}{2}\right),\quad \mbox{even\, m}.
\label{D12}
\end{gather} 
In spherical coordinates $k_x=k\sin\chi\cos\varphi$, $k_y=k\cos\chi$ and $k_z=k\sin\chi\sin\varphi$ the factor $\pi^2m^2-k_x^2L_x^2$ in the denominators of Eqs. (\ref{D11}) and (\ref{D12}) becomes $\pi^2m^2[1-\Gamma^{-2}\sin^2\chi\cos^2\varphi]$. Because  $k_x^2L_x^2\ll \pi^2m^2$ at $\Gamma^2
\gtrsim \epsilon_c\gg 1$, Eqs. (\ref{D11}) and (\ref{D12}) reduce to Eqs. (\ref{po}) and (\ref{pe}).

\end{document}